\documentclass[aps,pre,twocolumn,superscriptaddress,longbibliography]{revtex4-2}

\usepackage{amsmath,amssymb}
\usepackage{graphicx}
\usepackage{bm}
\usepackage{hyperref}
\usepackage{xcolor}

\newcommand{\avg}[1]{\langle #1 \rangle}
\newcommand{\betac}{\beta_c}
\newcommand{\psurviv}{P_{\rm surv}}
\newcommand{\fburn}{f_{\rm burn}}
\newcommand{\TFFM}{TFFM}
\newcommand{\WFFM}{WFFM}
\newcommand{\dw}{\delta_w}
\newcommand{\ew}{\hat{e}_w}
\newcommand{\eij}{\hat{e}_{ij}}

\begin{document}

\title{Topographic Disorder, Wind Coupling, and Directional Fire Spread:\\
       Critical Behavior in a Terrain-Weighted Forest Fire Model}

\author{Juan M. Florez}
\affiliation{Departamento de Ingeniería y Arquitectura, Universidad Central de Chile, Santiago, Chile}
\author{Eric Suárez Morell}
\affiliation{Grupo de Simulaciones, Departamento de Física, Universidad Técnica Federico Santa María, Valparaíso, Chile}
\author{Cristian Millan}
\affiliation{Departamento de Ingeniería y Arquitectura, Universidad Central de Chile, Santiago, Chile}
\author{J. Restrepo}
\affiliation{Grupo de Magnetismo y Simulación G+, Instituto de Física, Universidad de Antioquia, A.A. 1226, Medellín, Colombia}

\date{\today}

\begin{abstract}
We introduce the Terrain-Weighted Forest Fire Model (\TFFM), a lattice model
in which fire spreads on a spatially correlated Gaussian height field with the
asymmetric bond probability
$p_{i\to j}=\operatorname{clip}[e^{-\beta+\gamma(h_j-h_i)},0,1]$, plus an
additive wind bias of strength $\dw$ and direction $\theta$.  Simulations on
lattices up to $L=8192$ reveal a sharp active-to-inactive transition whose
critical suppression threshold $\betac$ is even in $\gamma$, decreases with
$|\gamma|$, and decreases as the terrain correlation length $\sigma_h$ is
reduced: slope asymmetry acts as a suppressant because downhill bonds are
penalized and fire stalls at local elevation maxima.  For rough terrain and
low tree density the fire fails to percolate even at zero suppression.
Finite-size scaling on a fine $\beta$ grid at $L=2048$--$8192$ gives a
correlation-length exponent $\nu=1.8\pm0.17$ from both the susceptibility
peak and the width of the transition, and a front-velocity exponent
$\delta=0.34\pm0.03$, identical for smooth and rough terrain; neither
matches directed percolation ($\nu_\perp=0.73$, $\nu_\parallel-\nu_\perp\approx0.56$)
or isotropic percolation ($\nu=4/3$, $\approx0.18$).  The single-seed
survival probability at $\betac$ is independent of $L$ and decays extremely
slowly, with a running exponent falling from $\approx0.09$ to $\approx0.04$,
excluding directed percolation and pointing to a survival probability that
remains finite at criticality, consistent with the $L$-independent value
$P^*\approx0.5$ at which $\psurviv$ drops to zero.  Wind raises $\betac$
by a factor of $2$--$4$, produces a sharp onset of downwind fire-scar drift
at weak coupling, and, at high terrain coupling, decreases the burned
fraction at boundary crossing---a terrain-wind competition effect absent
from isotropic bond-disorder models.  The model yields fire-risk thresholds
and fire-scar signatures directly comparable to satellite burn-scar data.
\end{abstract}


\maketitle

\section{Introduction}
\label{sec:intro}

Forest and wildland fires are among the most consequential natural disturbances
shaping ecosystem dynamics and pose an escalating risk to human
infrastructure~\cite{Sullivan2009,Bowman2017}.  Their spread is governed by an
interplay of fuel availability, atmospheric conditions, and
terrain~\cite{Rothermel1972,Finney1998,Sharples2016}.  Of these, terrain is
distinctive: it is spatially fixed, can be characterized with high precision
from digital elevation models, and exerts a systematic directional bias on
fire propagation---updrafts accelerate upslope spread while the geometry of the
heated gas column retards it on the opposite face~\cite{Sharples2009}.  Wind
is the other dominant directional driver: sustained downwind fire spread is the
primary factor distinguishing ordinary fires from catastrophic runs~\cite{Cruz2017,WindSlope2026}.
Yet, despite the importance of both topography and wind in operational fire
behavior~\cite{Finney1998}, their roles in the statistical physics of fire
percolation have not been systematically studied at the lattice-model level.

A concrete illustration of the practical stakes is provided by south-central
Chile, one of the world's leading exporters of pulp and timber.  Chile's
industrial forestry sector cultivates approximately 3 million ha of
\textit{Pinus radiata} and eucalyptus plantations~\cite{Bowman2019Chile,Lindenmayer2023}
---spatially uniform, high-fuel-continuity monoculture stands whose homogeneous
fuel structure is naturally approximated by a regular lattice model.
Although these plantations cover only $\sim$18\% of Chile's forested land, they
account for $\approx$30\% of the total area burned in each major fire
season~\cite{Bowman2019Chile}.
The 2017 fire season set Chile's modern record at over 580\,000 ha burned in a
single summer---14 times the 40-year mean~\cite{Bowman2019Chile};
the February~2024 Valpara\'{\i}so fire killed more than 130 people and
inflicted several billion dollars in damage despite burning only
$\sim$$7\,000$~ha~\cite{ChileDataset2025,CONAF2026}.
These events mirror a global trend toward more frequent and more severe
extreme fire seasons, driven by warming and intensifying fire
weather~\cite{Jones2024,Cordero2024,GlobalShifts2026}.
Percolation-based and stochastic lattice models~\cite{Rodrigues2021,LealMelo2022,Beneduci2024} offer a principled
theoretical framework for identifying critical fire-spread thresholds from first
principles, and the quantitative outputs derived here---the critical suppression
threshold $\betac(\gamma,\sigma_h)$ and the velocity exponent $\delta$ near
criticality---provide direct predictors for the suppression resources required to
keep plantation fires below their spreading threshold.

We ground this modeling framework in direct observational evidence.
Satellite analysis of Mediterranean fire scars~\cite{Caldarelli2001} reveals
fractal dimensions $D_f\approx1.90$--$1.95$, consistent with critical
percolation ($D_f=91/48\approx1.896$), and anisotropic propagation under
terrain and wind has been related to directed-percolation
universality~\cite{Porterie2008}.
The catastrophic February~2024 Valpara\'iso fire provides a concrete case study:
satellite imagery shows fire channeled through valley corridors under strong
winds and growing several-fold within hours~\cite{NASAValpo2024}---the
strongly channeled, wind-driven regime that our model reproduces at large
$\gamma$ and finite $\dw$.

Lattice models of fire spread have a long history as paradigmatic examples of
spreading phenomena, self-organized criticality, and
percolation~\cite{Stauffer1994,Saberi2015}.  The earliest forest-fire
self-organized criticality model~\cite{Bak1990,Jensen1998} was followed by the
Drossel--Schwabl model~\cite{Drossel1992}, which introduced tree growth and
lightning ignition and exhibits scale-free fire clusters.  More minimal models
connecting forest fires to directed percolation and contact
processes~\cite{Grassberger1993,Hinrichsen2000,Odor2004} have clarified the
universality class of the propagating front.  Recent work introduced quenched bond disorder
as a realistic mechanism for heterogeneous fire spread~\cite{wffm2025}: in the
Weighted Forest Fire Model (\WFFM), each bond carries a random weight with a
suppression parameter $\beta$ controlling the overall spreading while preserving
a spatially heterogeneous structure.

Despite its success in reproducing the qualitative features of the
active-to-inactive transition, the \WFFM\ has no spatial structure: its bond
weights are uncorrelated and isotropic, and wind is absent.  Cellular automaton
fire models have incorporated wind direction as an empirical spread-rate
multiplier~\cite{Alexandridis2008}, but without the percolation-based threshold
framework or systematic terrain coupling developed here.  Real landscapes,
by contrast, have spatially correlated height fields---hills, valleys, and
ridgelines spanning length scales from meters to kilometers---and fire responds
to these correlations directionally.  Capturing this physics requires a model in
which (i)~the bond disorder has a tunable spatial correlation length,
(ii)~the spreading probability is asymmetric ($p_{i\to j}\neq p_{j\to i}$)
whenever there is a height difference, and (iii)~a wind term can be added as an
independent directional bias.

Here we introduce the Terrain-Weighted Forest Fire Model (\TFFM) and extend it
to include wind coupling.  The terrain is modeled as a spatially correlated
Gaussian random field with correlation length $\sigma_h$, and fire spreads with
a slope-biased probability that depends on the local height difference through a
coupling parameter $\gamma$.  Setting $\gamma=0$ recovers a disorder-free
spreading model.  Wind enters as an additive term in the exponent, parameterized
by coupling strength $\dw$ and direction $\theta$.

Our principal contributions are as follows.
We provide a systematic mapping of the phase boundary $\betac(\gamma,\sigma_h,p)$
across all four tree densities $p\in\{0.7,0.8,0.9,1.0\}$, quantifying how
terrain coupling and terrain roughness jointly control the onset of large-scale
fire propagation.
We show that $\betac(\sigma_h)$ decreases monotonically as $\sigma_h$
decreases: rough terrain, in which neighboring sites differ in height by
$\mathcal{O}(1)$ standard deviations, maximally hinders fire percolation
because the penalized downhill bonds trap the fire at local maxima; for
sufficiently small $\sigma_h$ and $p$, $\betac$ falls below zero and the
fire fails to percolate even at zero suppression.
We introduce the fire-shape anisotropy $\eta$---derived from the inertia tensor
of the burned cluster---as a directly observable signature of terrain-driven
directional fire spread.
We perform a finite-size scaling analysis at $L\in\{256,\dots,8192\}$ for
both $\sigma_h=10$ and $\sigma_h=1$, including a fine $\beta$ grid
($\Delta\beta=0.001$) at the three largest sizes, which yields the
thermodynamic thresholds $\betac^\infty$, a correlation-length exponent
$\nu=1.8\pm0.2$ from two independent estimators, and a velocity exponent
$\delta=0.34\pm0.03$; both exponents are the same for smooth and rough
terrain and neither matches isotropic or directed percolation.  We complement
the static analysis with the time-dependent survival probability $P(t)$ of a
single-seed fire at $\betac^\infty$, which excludes directed percolation
directly and points to a finite survival probability at criticality.  We also
show that coarse $\beta$ grids produce spurious, erratic estimates of $\nu$
in this model, a methodological point relevant to other quenched-disorder
spreading models~\cite{Weinrib1983,Vojta2006}.
Finally, we present a wind coupling extension revealing a monotone $\betac(\dw)$
shift, a clipping-saturation regime for axial wind, a wind-direction drift with
sharp onset at weak coupling, and---crucially---a terrain-wind competition regime
in which sufficiently strong $\gamma$ causes the burned fraction at boundary
crossing to \emph{decrease} with east-wind intensity, an inversion absent in
isotropic bond-disorder models.

The paper is organized as follows.  Section~\ref{sec:model} defines the \TFFM\
and its wind extension.  Section~\ref{sec:obs} specifies the observables and
the finite-size scaling approach.  Section~\ref{sec:results} presents the
simulation results, including terrain phase diagrams (Secs.~\ref{sec:results_gamma}--\ref{sec:results_phase}),
finite-size scaling (Sec.~\ref{sec:results_fss}), the wind sweep
(Sec.~\ref{sec:results_wind}), and fire-scar morphology
(Sec.~\ref{sec:results_morph}).  Section~\ref{sec:discussion} discusses the
results in the context of the \WFFM, universality, and wildfire science.
Section~\ref{sec:conclusions} summarizes our conclusions.

\begin{figure*}[t]
\includegraphics[width=2.0\columnwidth]{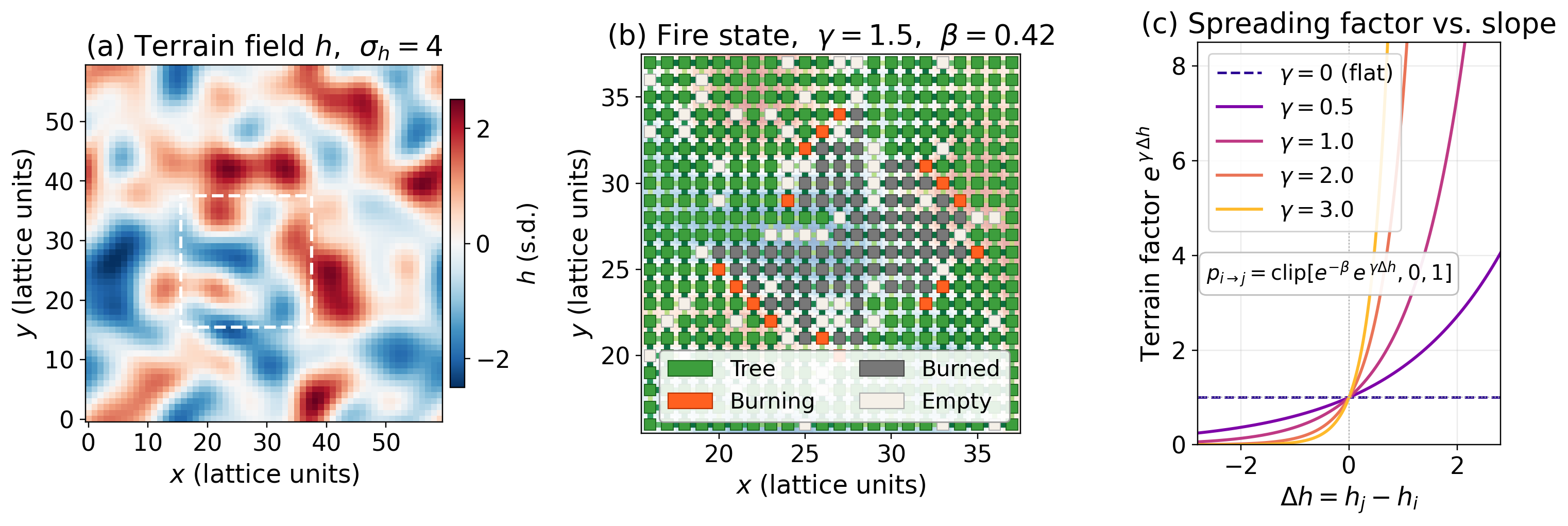}
\caption{%
  Schematic of the \TFFM.
  (a)~Height field $h$ with correlation length $\sigma_h$ (cool = low, warm = high).
  (b)~Fire state: burning (orange), burned (gray), unburned tree (green).
  Bond width indicates $p_{i\to j}$; uphill bonds are enhanced for $\gamma>0$.
  (c)~Terrain factor $e^{\gamma\Delta h}$ vs $\Delta h=h_j-h_i$;
  $\gamma=0$ (dashed) gives uniform spreading at $e^{-\beta}$.
}
\label{fig:schematic}
\end{figure*}

\section{Model and methods}
\label{sec:model}

\subsection{Lattice, states, and initial conditions}

We consider an $L\times L$ square lattice $\Lambda$ with open (absorbing)
boundaries.  Each site $i\in\Lambda$ occupies one of four states: empty~(0),
tree~(1), burned~(2), or burning~(3).  Trees are distributed independently:
each site is a tree with probability $p$ and empty with probability $1-p$.
We study $p\in\{0.7,0.8,0.9,1.0\}$, all above the 2D site-percolation
threshold $p_c\approx 0.593$.  The central site $(L/2,L/2)$ is always a tree
and is set to the burning state at $t=0$; all other trees start unburned.

\subsection{Terrain generation}
\label{sec:terrain}

A height field $h:\Lambda\to\mathbb{R}$ is drawn independently for each
realization (quenched disorder).  Generation proceeds in three steps:

(1) Draw i.i.d.\ Gaussian noise $z_{ij}\sim\mathcal{N}(0,1)$ for all
  $(i,j)\in\Lambda$.

(2) Apply a Gaussian low-pass filter in Fourier space,
  \begin{equation}
    \hat{h}(\mathbf{k}) = \hat{z}(\mathbf{k})\,
    \exp\!\bigl(-2\pi^2\sigma_h^2\,|\mathbf{k}|^2\bigr),
    \label{eq:fourier}
  \end{equation}
  and set $h=\operatorname{Re}[\mathcal{F}^{-1}\hat{h}]$.  The parameter
  $\sigma_h$ (in lattice spacings) controls the spatial correlation length:
  $\sigma_h=0$ gives uncorrelated site heights, while large $\sigma_h$
  produces slowly varying, smooth landscapes.
  
(3) Normalize: $h\leftarrow(h-\bar{h})/s_h$, where $\bar{h}$ and $s_h$
  are the sample mean and standard deviation of the filtered field, so that
  the terrain amplitude is $\mathcal{O}(1)$ regardless of $\sigma_h$.
  Because the amplitude is fixed, $\sigma_h$ controls the typical
  nearest-neighbor height difference: $|h_j-h_i|$ is $\mathcal{O}(1)$ for
  $\sigma_h\lesssim1$ and $\mathcal{O}(1/\sigma_h)$ for $\sigma_h\gg1$.

\subsection{Spreading rule}
\label{sec:spread}

Fire at site $i$ attempts to ignite each tree neighbor $j$ independently with
probability
\begin{equation}
  p_{i\to j}(\beta,\gamma)
  = \operatorname{clip}\!\bigl[\,e^{-\beta}\,e^{\,\gamma(h_j-h_i)},\;0,\;1\,\bigr].
  \label{eq:spread}
\end{equation}
The update is synchronous: at each discrete time step all currently burning
sites attempt to spread simultaneously, after which all burning sites become
burned and all newly ignited sites become burning.  Each step therefore
corresponds to one fire-generation interval; the equal-arrival-time isochrones
visible in the scar panels of Figs.~\ref{fig:app_morph} and~\ref{fig:app_paradox} directly display
this temporal structure, and physical elapsed time is recovered by mapping one
step to a known fire-spread rate (e.g., meters per minute for a given fuel type),
making the model straightforwardly calibratable to field observations.

Three aspects of Eq.~(\ref{eq:spread}) are worth noting.  (i)~$e^{-\beta}$ is
the base spreading probability; $\beta=0$ gives $p_{i\to j}=1$ (maximum
spreading) while $\beta\to\infty$ extinguishes fire.  (ii)~$e^{\gamma(h_j-h_i)}$
is the terrain factor; for $\gamma>0$, bonds from lower to higher sites receive
enhanced spreading probability, which is the physically realistic case of
preferential upslope spread~\cite{Sharples2009}; the opposite convention
$\gamma<0$ favors downhill spread.  Since the Gaussian terrain is
statistically symmetric under $h\to -h$, all results satisfy
$\betac(\gamma)=\betac(-\gamma)$ to numerical precision, and we quote
results for $\gamma>0$ without loss of generality.
(iii)~The clip to $[0,1]$ enforces that $p_{i\to j}$ is a valid probability.

Because $h_j-h_i=-(h_i-h_j)$, the spreading probability is directional:
$p_{i\to j}\neq p_{j\to i}$ whenever $h_i\neq h_j$, breaking the bond
symmetry present in the \WFFM.

The \TFFM\ has four control parameters: $\beta$ (suppression), $\gamma$
(terrain coupling), $\sigma_h$ (terrain correlation length), and $p$
(tree density).  Setting $\gamma=0$ recovers a uniform spreading model.
Figure~\ref{fig:schematic} illustrates the model geometry.

\subsection{Wind coupling}
\label{sec:wind}

We extend the spreading rule to include a wind directional bias.  For a wind
blowing in direction $\theta$ (measured from east), the spreading probability
becomes
\begin{multline}
  p_{i\to j}(\beta,\gamma,\dw,\theta)
  = \operatorname{clip}\!\Bigl[\exp\!\bigl(-\beta \\
    +\gamma(h_j-h_i)+\dw(\ew\cdot\eij)\bigr),\;0,\;1\Bigr],
  \label{eq:spread_wind}
\end{multline}
where $\dw\geq 0$ is the wind coupling strength, $\ew=(\cos\theta,\sin\theta)$
is the unit vector in the wind direction, and $\eij$ is the unit vector from
site $i$ to site $j$.  On the square lattice the four bond projections are:
eastward $+\cos\theta$, westward $-\cos\theta$, southward $+\sin\theta$,
northward $-\sin\theta$.  Setting $\dw=0$ recovers the terrain-only
Eq.~(\ref{eq:spread}).

The wind term couples additively to the terrain factor in the exponent.
A bond aligned with the wind direction ($\ew\cdot\eij>0$) receives enhanced
spreading probability, while an opposing bond is suppressed by the same amount.
At large $\dw$, downwind bonds are clipped to probability~1 and upwind bonds to
probability~0, effectively reducing fire to a directed process in the wind
direction.  We study $\dw\in\{0,0.25,0.50,0.75,1.00,1.50,2.00\}$ at two wind
directions: $\theta=0^{\circ}$ (axial, east) and $\theta=45^{\circ}$ (diagonal,
northeast), at $\gamma=1.0$, $\sigma_h=10$, $p\in\{0.8,1.0\}$.

\subsection{Relationship to the WFFM}
\label{sec:relation}

The \WFFM~\cite{wffm2025} uses symmetric, spatially uncorrelated bond weights
$w_{ij}=e^{\beta(r_{ij}-1)}$ with $r_{ij}\sim U[0,1]$.  The \TFFM\ (with or
without wind) differs in three key respects: (i)~spreading is asymmetric
($p_{i\to j}\neq p_{j\to i}$); (ii)~bond disorder is spatially correlated over
length $\sigma_h$; and (iii)~the active phase is sustained by coherent terrain
pathways rather than rare individual bonds.  The \WFFM\ critical threshold at
$p=0.8$ is $\betac\approx 1.01$~\cite{wffm2025}, considerably larger than the
\TFFM\ values reported here, reflecting the qualitatively different spreading
mechanisms.

\section{Observables and methods}
\label{sec:obs}

We compute five observables per $(\beta,\gamma,\sigma_h,p)$ point, each
averaged over $N$ independent realizations.  For wind runs, a sixth observable
(wind-direction drift) is added.

\subsection{Average fire-front velocity}
\begin{equation}
  v = \frac{1}{t_{\rm hit}}\,\overline{d(\text{burning sites, center})}
  \label{eq:vel}
\end{equation}
for realizations that reach the boundary at time $t_{\rm hit}$; $v=0$
for realizations in which fire extinguishes before reaching the boundary.

\subsection{Fire survival probability and susceptibility}
\label{sec:psurviv_def}
\begin{equation}
  \psurviv(\beta) = \frac{1}{N}\sum_{k=1}^{N}
  \mathbf{1}[\text{realization }k\text{ reaches boundary}].
\end{equation}
$\psurviv\to 1$ in the active phase and $\psurviv\to 0$ in the inactive phase
as $L\to\infty$.  We use the slope
$\chi(\beta)=|d\psurviv/d\beta|$
as a susceptibility-like quantity: it peaks sharply at $\betac$, provides an
independent estimate of the critical point, and its peak height scales with
the inverse width of the finite-size transition region.

\subsection{Mean burned fraction}
\begin{equation}
  \fburn = \frac{|\{i : s_i \geq 2\}|}{|\{i : s_i \geq 1\}|},
\end{equation}
which is the ratio of burned to total tree sites at the end of the simulation.
Unlike $\avg{v}$, which is zero for non-percolating fires, $\avg{\fburn}$
captures damage from both percolating and non-percolating events and
is the natural fire-damage metric for risk assessment.

\subsection{Shape anisotropy}
\label{sec:eta_def}

Let $\{(x_i,y_i)\}$ denote the positions of burned sites $(s_i\geq 2)$ relative
to the ignition center.  The $2\times 2$ inertia tensor is
\begin{equation}
  \mathbf{I} = \begin{pmatrix}\avg{x^2} & \avg{xy}\\\avg{xy} & \avg{y^2}\end{pmatrix}
\end{equation}
with eigenvalues $\lambda_1\geq\lambda_2\geq 0$.  The shape anisotropy index is
\begin{equation}
  \eta = \frac{\lambda_1 - \lambda_2}{\lambda_1}\;\in\;[0,1].
  \label{eq:eta}
\end{equation}
A circular burned cluster gives $\eta=0$; a maximally elongated (linear)
cluster gives $\eta\to 1$.  For a uniformly filled ellipse with semi-axes
$a\geq b$, $\eta=1-(b/a)^2$.

\subsection{Normalized fire-front roughness}

At time $t_{\rm hit}$, let $d_i$ be the distance of burning site $i$ from the
ignition center.  The normalized roughness is
\begin{align}
  W &= \frac{\sigma_d}{\bar{d}},\quad \bar{d} = \avg{d_i}, \notag\\
  \sigma_d &= \sqrt{\avg{d_i^2}-\avg{d_i}^2}.
  \label{eq:rough}
\end{align}
$\avg{W}$ is averaged over surviving realizations only.

\subsection{Wind-direction drift}
\label{sec:drift_def}

For wind simulations, we define the center-of-mass displacement of the burned
cluster projected onto the wind direction:
\begin{equation}
  d_w = (\bar{x}_{\rm burn} - x_0)\cos\theta
      + (\bar{y}_{\rm burn} - y_0)\sin\theta,
  \label{eq:drift}
\end{equation}
where $(\bar{x}_{\rm burn},\bar{y}_{\rm burn})$ is the mean position of all
burned sites and $(x_0,y_0)$ is the ignition site.  Positive $d_w$ indicates
systematic downwind displacement of the fire scar center of mass.  By symmetry,
$\avg{d_w}=0$ at $\dw=0$.

\subsection{Time-dependent survival probability}
\label{sec:pt_def}

To probe the dynamics at criticality directly, we also record, for fires
started from a single site at $\beta=\betac^\infty$, the fraction $P(t)$ of
realizations that still contain at least one burning site after $t$ steps.
Realizations that reach the absorbing boundary at a time $t_{\rm hit}$ are
right-censored (Kaplan--Meier estimator), so that $P(t)$ is unbiased for
$t<\min t_{\rm hit}$, which grows linearly with $L$.  In directed percolation
$P(t)\sim t^{-\delta_{\rm DP}}$ with $\delta_{\rm DP}=0.4505$ in $(2{+}1)$
dimensions~\cite{Hinrichsen2000}; in isotropic percolation with synchronous
burning dynamics the probability that growth from a seed reaches chemical
distance $t$ scales as $t^{-\beta_{\rm exp}/(\nu d_{\min})}\approx t^{-0.092}$
in two dimensions~\cite{Stauffer1994,Grassberger1992}; at an
infinite-randomness fixed point $P(t)$ decays as a power of
$\ln t$~\cite{Vojta2005}.

\subsection{Finite-size scaling}
\label{sec:fss}

Near $\betac$, standard finite-size scaling~\cite{Henkel2008} predicts
\begin{align}
  \avg{v}    &\sim |\beta-\betac|^{\delta}\,
               f_v\!\bigl(L^{1/\nu}(\beta-\betac)\bigr), \label{eq:fss_v}\\
  \psurviv   &\sim |\beta-\betac|^{\beta_{\rm exp}}\,
               g\!\bigl(L^{1/\nu}(\beta-\betac)\bigr), \label{eq:fss_p}\\
  \betac(L)  &= \betac + a\,L^{-1/\nu}. \label{eq:fss_bc}
\end{align}
We extract $\betac$ and the exponents $\nu$, $\delta$, $\beta_{\rm exp}$ from
runs at $L\in\{256,512,1024,2048,4096,8192\}$.

\subsection{Critical-point extraction}
\label{sec:betac_method}

We extract $\betac$ by two complementary methods.  \textit{$\psurviv=0.5$
crossing:} linear interpolation of the monotone $\psurviv(\beta)$ at the
crossing value, which is robust against finite-$N$ noise.  \textit{Susceptibility
peak:} the peak of $\chi(\beta)=|d\psurviv/d\beta|$, computed by centered
finite differences, either on the raw $\psurviv(\beta)$ data or after
Gaussian smoothing over $1.5$ grid spacings.
At $L=2048$ the two estimates agree to within $\approx0.01$ in $\betac$; the
susceptibility-peak estimate lies systematically above the crossing estimate
at smaller $L$ (cf.\ the legend of Fig.~\ref{fig:fss}).
The $\psurviv=0.5$ crossing is the primary estimate throughout.  For the
finite-size scaling analysis, the thermodynamic limit $\betac^\infty$ is
estimated by the \textit{pairwise-crossing method}: the value of $\beta$ at
which $\psurviv(L_1,\beta)=\psurviv(L_2,\beta)$ in the transition region
$\psurviv\in(0.25,0.75)$.  Because $\psurviv$ at the crossing converges
toward a universal value $P^*$ as both sizes grow, we take the crossing of the
two largest sizes as the estimate of $\betac^\infty$ and use the sequence of
consecutive-pair crossings to gauge its convergence.

\subsection{Simulation parameters}
\label{sec:simparams}

All bulk results use $L=2048$ with $N=2000$ independent realizations per
$(\beta,\gamma,\sigma_h,p)$ point.  Finite-size scaling uses
$L\in\{256,512,1024,2048\}$ with $N=2000$--$3000$ at each size, supplemented
by targeted runs at $L=4096$ and $L=8192$ with $N=1500$, and by fine-grid runs
at $L=2048$, $4096$, and $8192$ ($\Delta\beta=0.001$ over
$\betac^\infty\pm0.03$ plus eight flanking points; $N=3000$, $3000$, $1500$),
which supersede the coarse-grid data at those sizes in the finite-size
scaling analysis.  The survival probability $P(t)$ is measured at $\betac^\infty$
for $L\in\{2048,4096,8192\}$ with $N=3000$, $3000$, $1500$.  Wind sweep
runs use $L=2048$, $N=2000$, $\gamma=1.0$, $\sigma_h=10$, $p\in\{0.8,1.0\}$,
covering 13 wind configurations
($\dw\in\{0,0.25,0.50,0.75,1.00,1.50,2.00\}$, $\theta\in\{0^{\circ},45^{\circ}\}$
plus the $\dw=0$ baseline).

Each parameter set uses an adaptive three-zone $\beta$ grid totaling
$\approx 60$--$80$ values, with 40 fine points concentrated within
$\betac\pm0.10$ and coarser sampling away from the transition.

Fire propagation is implemented as synchronous breadth-first search.  The terrain
field is generated once per realization via FFT [Eq.~(\ref{eq:fourier})] and
reused across all $\beta$ values in a loop-inversion strategy that amortizes the
terrain generation cost.
The final full parameter sweep (Phases~A through F) required approximately
$80$--$90$~CPU-hours (10-16 cores); the wind sweep required a further $\approx 38$~CPU-hours;
the FSS targeted runs at $L=4096$ together with the $\sigma_h=3$ survey
required a further $\approx 55$~CPU-hours; the targeted $L=8192$ FSS run
at $\sigma_h=10$ required a further $\approx 92$~CPU-hours; and the targeted
$L=8192$ FSS run at $\sigma_h=1$ required a further $\approx 55$~CPU-hours,
the fine-grid runs at $L=2048$--$8192$ required $\approx 650$~CPU-hours and
the survival-probability runs $\approx 60$~CPU-hours, bringing the total
project's displayed-data compute to approximately $1000$~CPU-hours.
Wall time scales as $L^2$ per realization at fixed $N$: doubling $L$ quadruples
the computational cost, so the two $L=8192$ runs together account for the
dominant share of the FSS budget.

\section{Results}
\label{sec:results}

All results use $L=2048$ unless otherwise stated, verified to be in the
thermodynamic regime by the FSS analysis of Sec.~\ref{sec:results_fss}.
Each data point represents $N=2000$ realizations except for the FSS study
($N=2000$--$3000$ at each $L$).  Error bars on $\betac$ reflect the uncertainty
from the $\psurviv=0.5$ interpolation, typically $\pm 0.003$--$0.005$.

\subsection{Phase transition: all observables at varying terrain coupling}
\label{sec:results_gamma}

\begin{figure*}[t]
\includegraphics[width=\textwidth]{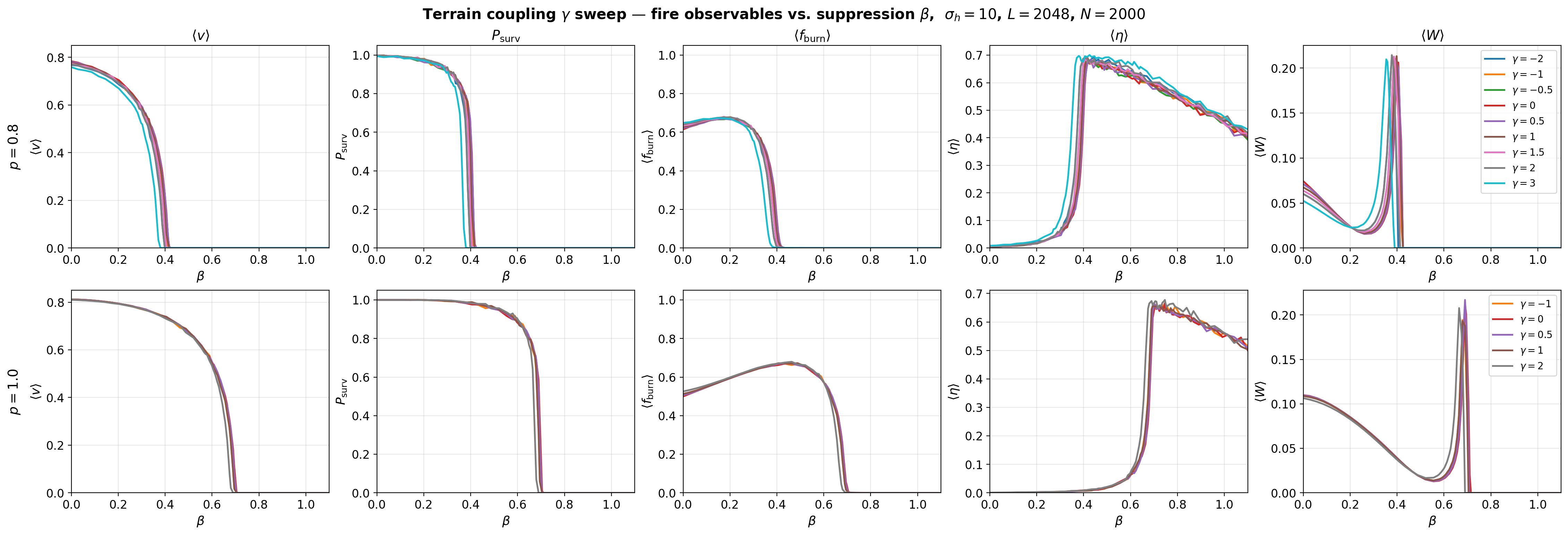}
\caption{%
  All five observables vs $\beta$ for
  $\gamma\in\{-2,-1,-0.5,0,0.5,1,1.5,2,3\}$ at $p=0.8$ (top row) and
  $\gamma\in\{-1,0,0.5,1,2\}$ at $p=1.0$ (bottom row), at $\sigma_h=10$,
  $L=2048$, $N=2000$;
  columns: $\avg{v}$, $\psurviv$, $\avg{\fburn}$, $\avg{\eta}$, $\avg{W}$.
  All observables show a sharp transition at $\betac(\gamma)$, which
  decreases monotonically with $|\gamma|$ (most visibly for $\gamma=3$).
  $\avg{\eta}$ is small in the deep active phase, rises steeply near
  $\betac$, and remains large in the inactive phase, where the small
  contained clusters are strongly anisotropic.
}
\label{fig:phaseA}
\end{figure*}

Figure~\ref{fig:phaseA} shows all five observables as a function of $\beta$
for terrain coupling $\gamma\in[-2,3]$ at
$\sigma_h=10$, for tree densities $p=0.8$ (upper row) and $p=1.0$ (lower row).

In all cases $\avg{v}$ is large in the active phase ($\beta < \betac$) and
drops sharply to zero at the critical threshold.  The transition sharpens
with larger $p$, reflecting the fact that denser forests provide more reliable
spreading pathways.  Increasing $|\gamma|$ shifts $\betac$ to \emph{smaller}
values: the slope asymmetry of Eq.~(\ref{eq:spread}) enhances uphill bonds
but penalizes downhill ones by the same factor, and because the uphill gain
is capped by the clip at unity while the downhill loss is not, the net effect
of terrain coupling is to reduce the connectivity of the spreading network.
Fire that has climbed to a local elevation maximum finds all outgoing bonds
suppressed, so ridgelines act as natural firebreaks.

The fire survival probability $\psurviv$ displays a sharp sigmoid transition
at the same $\betac$ as $\avg{v}$, confirming that both observables locate the
same critical point.  At $\gamma=0$ (terrain-free baseline) we find
$\betac=0.405\pm0.003$ for $p=0.8$ and $\betac=0.692\pm0.003$ for $p=1.0$.
Over the range $|\gamma|\in[0,3]$, $\betac$ falls from $0.405$ to $0.361$
($p=0.8$), a reduction of $\sim\!11\%$; for $p=1.0$ at $\gamma=2.0$,
$\betac$ falls from $0.692$ to $0.668$ ($\sim\!3.5\%$), with the largest
value $\gamma=3$ not simulated for $p=1.0$
(Table~\ref{tab:betac_gamma}).  The proportional reduction grows as the tree
density decreases ($\sim\!7\%$ at $p=0.9$, $\sim\!23\%$ at $p=0.7$),
indicating that terrain coupling has the strongest leverage when forest
connectivity is itself marginal.

The mean burned fraction $\avg{\fburn}$ approaches unity in the deep active
phase---virtually every fire percolates---and falls through the transition to
a residual contribution from non-percolating events in the inactive phase.
It is worth noting that $\avg{\fburn}$ is nonzero even at $\beta>\betac$
because small contained fires still burn a finite fraction of trees.
Because $\betac$ moves to lower $\beta$ with $|\gamma|$, a suppression level
that is marginally sub-critical on flat terrain ($\beta\approx0.40$ at
$p=0.8$) is already super-critical for $\gamma=3$, where $\avg{\fburn}$ has
collapsed to its residual value: at fixed management effort, strong slope
coupling reduces the expected damage in this model.

The shape anisotropy $\avg{\eta}$ is small ($\lesssim0.02$) in the deep
active phase for all $\gamma$, where the percolating scar is statistically
circular, rises steeply as $\beta\to\betac$, and reaches
$\avg{\eta}\approx0.7$ just above $\betac$ before decaying slowly at larger
$\beta$, where the surviving clusters are small and their shape is dominated
by the elongated, branched geometry of near-critical clusters.  At
$\sigma_h=10$ the $\gamma$ dependence of $\avg{\eta}$ is weak: the curves
for different $\gamma$ are shifted along $\beta$ together with $\betac$ but
have nearly the same amplitude.  The terrain contribution to the anisotropy
is instead controlled by $\sigma_h$ (Sec.~\ref{sec:results_sigmah}) and by
wind (Sec.~\ref{sec:results_wind}).

The normalized fire-front roughness $\avg{W}$ is small in the deep active phase
and grows sharply as $\beta$ approaches $\betac$ from below.  Both $\gamma$ and
$\sigma_h$ amplify $\avg{W}$ in the near-critical regime.

\subsection{Effect of terrain correlation length}
\label{sec:results_sigmah}

\begin{figure*}[t]
\includegraphics[width=\textwidth]{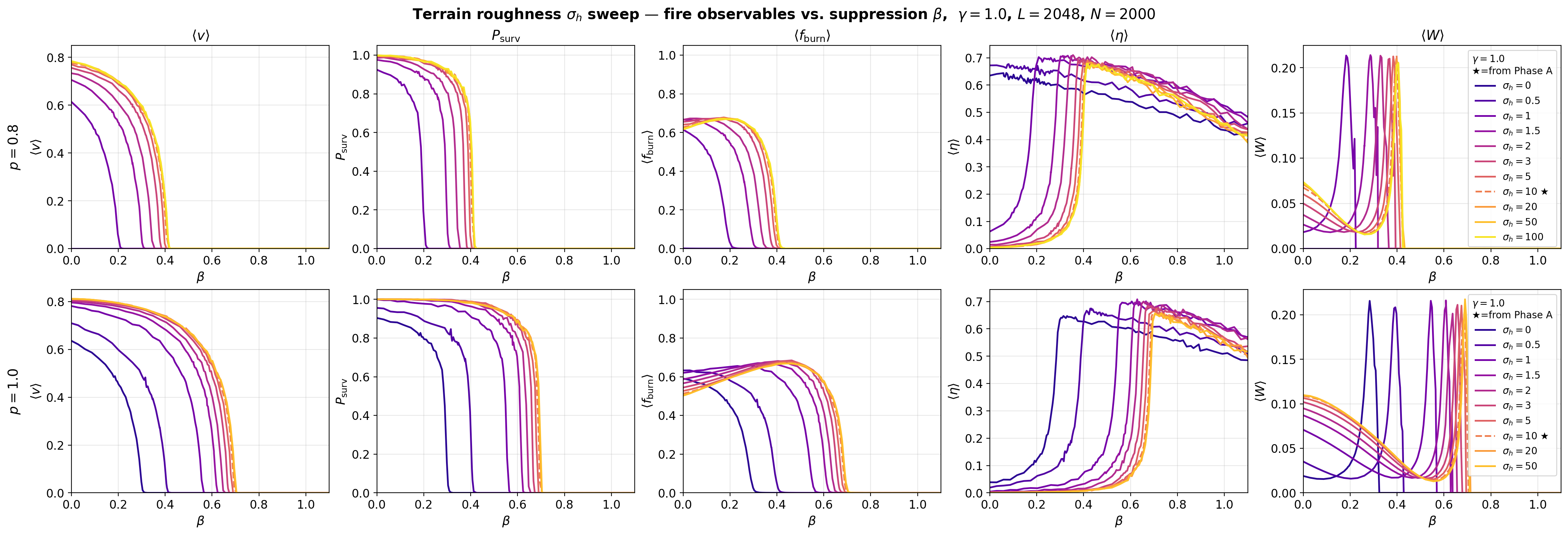}
\caption{%
  All five observables vs $\beta$ for
  $\sigma_h\in\{0,0.5,1,1.5,2,3,5,10,20,50,100\}$ (colored lines)
  at $\gamma=1.0$, $L=2048$, $N=2000$; rows: $p=0.8$ (top), $p=1.0$ (bottom).
  Dashed curve ($\sigma_h=10$, $\star$): Phase~A reference.
  $\betac$ decreases monotonically with decreasing $\sigma_h$; curves with
  $\psurviv=0$ throughout $\beta\in[0,1.1]$ (e.g., $\sigma_h=0$ at
  $p=0.8$) indicate $\betac<0$, i.e., fire does not percolate even at
  $\beta=0$.
  The near-critical rise of $\avg{\eta}$ is steeper and higher for large
  $\sigma_h$: smooth terrain sustains directional channels absent in rough
  terrain.
}
\label{fig:phaseB}
\end{figure*}

Figure~\ref{fig:phaseB} shows all five observables as a function of $\beta$
for terrain correlation lengths
$\sigma_h\in\{0,0.5,1,1.5,2,3,5,10,20,50,100\}$ at $\gamma=1.0$,
for $p=0.8$ (upper row) and $p=1.0$ (lower row).

Of particular interest is the monotone dependence of $\betac$ on
$\sigma_h$: decreasing $\sigma_h$ from the smooth-terrain plateau
($\sigma_h=100$, $\betac\approx 0.405$ at $p=0.8$) uniformly \emph{lowers}
$\betac$.  At $\sigma_h=1$, $\betac=0.189$---less than half the smooth-terrain
value.  For $\sigma_h\leq 0.5$ at $p=0.8$, the $\psurviv=0.5$ crossing is no
longer found in $\beta\in[0,1.1]$: $\psurviv$, $\avg{v}$, and $\avg{\fburn}$
are zero for all $\beta\geq0$, i.e., $\betac<0$ and the fire fails to
percolate even in the absence of suppression.  The same monotone trend
holds for $p=1.0$: $\betac(50)\approx 0.691$ (the smooth-terrain plateau;
$\sigma_h=100$ was not bracketed within the scanned range for $p=1.0$,
see Table~\ref{tab:betac_sigmah}), falling to $\betac(1)=0.547$,
$\betac(0.5)=0.394$, and reaching the white-noise minimum at
$\betac(0)=0.287$.  For lower tree densities $p=0.7$ and $p=0.8$, $\betac$
enters the always-inactive ($\betac<0$) regime already at larger $\sigma_h$;
Table~\ref{tab:betac_sigmah} marks those entries with ellipses.

The near-critical shape anisotropy $\avg{\eta}$ increases with $\sigma_h$,
reaching its largest values for smooth terrain ($\sigma_h\gtrsim 10$), while
for $\sigma_h\lesssim1$ the plateau above $\betac$ is lower and the rise is
broader.  This reflects the different spatial character of the
terrain: at small $\sigma_h$, the terrain is maximally heterogeneous on the
bond scale, but local anisotropies cancel on large scales.  For large $\sigma_h$,
the terrain is slowly varying, creating coherent gradient channels that span
many lattice spacings and sustain directional elongation of the fire scar.

The roughness $\avg{W}$ shows a similar pattern: large correlation lengths
produce more irregular fire fronts near $\betac$, while short-range correlated
terrain produces more symmetric critical clusters.

\subsection{Phase diagram and multi-density results}
\label{sec:results_phase}

\begin{figure*}[t]
\includegraphics[width=2.0\columnwidth]{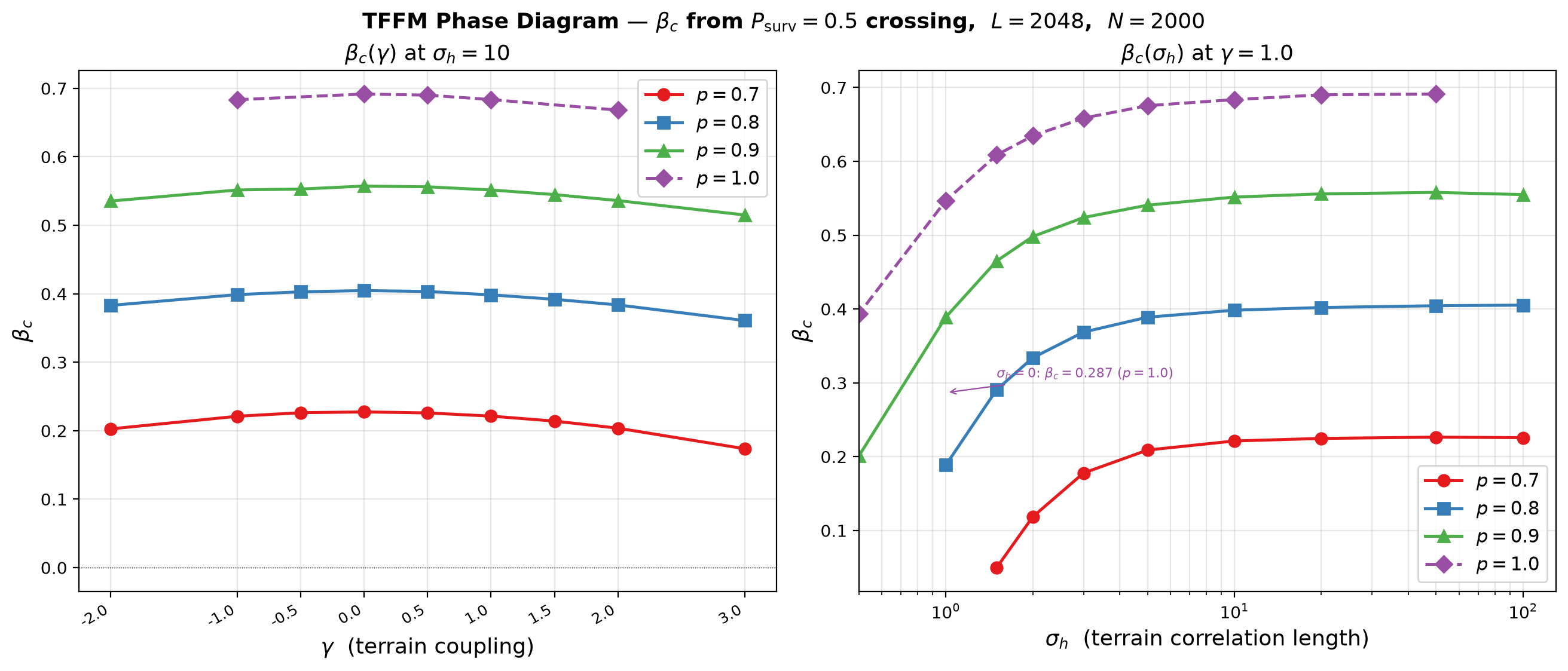}
\caption{%
  Phase diagram of the \TFFM\ ($\psurviv=0.5$ crossing, $L=2048$).
  (a)~$\betac$ vs $\gamma$ at $\sigma_h=10$ for $p\in\{0.7,0.8,0.9,1.0\}$:
  decreases monotonically with $|\gamma|$ and is approximately even-symmetric,
  $\betac(\gamma)\approx\betac(-\gamma)$ (Table~\ref{tab:betac_gamma}).
  (b)~$\betac$ vs $\sigma_h$ (log scale) at $\gamma=1.0$: decreases monotonically
  as $\sigma_h$ decreases; points with $\betac<0$ are omitted
  (Table~\ref{tab:betac_sigmah}).  The arrow marks the $\sigma_h=0$
  (white-noise) value for $p=1.0$.
}
\label{fig:phase}
\end{figure*}

Figure~\ref{fig:phase} shows the phase boundary $\betac$ as a function of
$\gamma$ (panel~a) and $\sigma_h$ (panel~b) for all four tree densities
$p\in\{0.7,0.8,0.9,1.0\}$.

\textit{$\gamma$ dependence.}  Table~\ref{tab:betac_gamma} summarizes the
measured critical thresholds at $\sigma_h=10$.  For all tree densities,
$\betac$ decreases monotonically with $|\gamma|$ and satisfies
$\betac(\gamma)\approx\betac(-\gamma)$ to within $\pm 0.001$.
This symmetry follows from the zero-mean symmetry of the Gaussian terrain:
reversing the sign of $\gamma$ maps the distribution of
$p_{i\to j}$ onto itself via $h\to -h$, leaving the percolation
properties unchanged.

Over the range $|\gamma|\in[0,3]$, $\betac$ decreases by approximately
$11\%$ for $p=0.8$ (from $0.405$ to $0.361$, Table~\ref{tab:betac_gamma}).
For $p=1.0$, the largest simulated $\gamma$ is $2.0$, giving a
$\sim\!3.5\%$ reduction (from $0.692$ to $0.668$).  The relative rate of
decrease grows monotonically as $p$ is lowered, reaching $\sim\!23\%$ at
$p=0.7$: the fewer the spreading pathways provided by tree connectivity, the
more effectively slope asymmetry removes them.

\textit{$\sigma_h$ dependence.}  Table~\ref{tab:betac_sigmah} summarizes
$\betac(\sigma_h)$ at $\gamma=1.0$ for all four $p$ values.
$\betac$ decreases monotonically as $\sigma_h$ decreases, for all tree
densities studied.  The decrease from the smooth-terrain plateau to $\sigma_h=1$,
$\Delta\betac = \betac(\sigma_h\to\infty) - \betac(1)$,
is $\approx0.14$ at $p=1.0$, $\approx0.17$ at $p=0.9$, and $\approx0.22$ at
$p=0.8$; at $p=0.7$ the threshold has already dropped below zero.

The saturation of $\betac$ at large $\sigma_h$ confirms the smooth-terrain
limit: $\betac(\sigma_h=50)\approx\betac(\sigma_h=100)$ for all $p$.
For $\sigma_h=0$ (white noise), the terrain factor fluctuates maximally from
bond to bond with no spatial coherence, yielding the lowest $\betac$ for $p=1.0$
($\betac(0)=0.287$) and pushing $\betac$ below zero for lower densities,
where site dilution and slope asymmetry together disconnect the spreading
network.

\begin{figure*}[t]
\includegraphics[width=\textwidth]{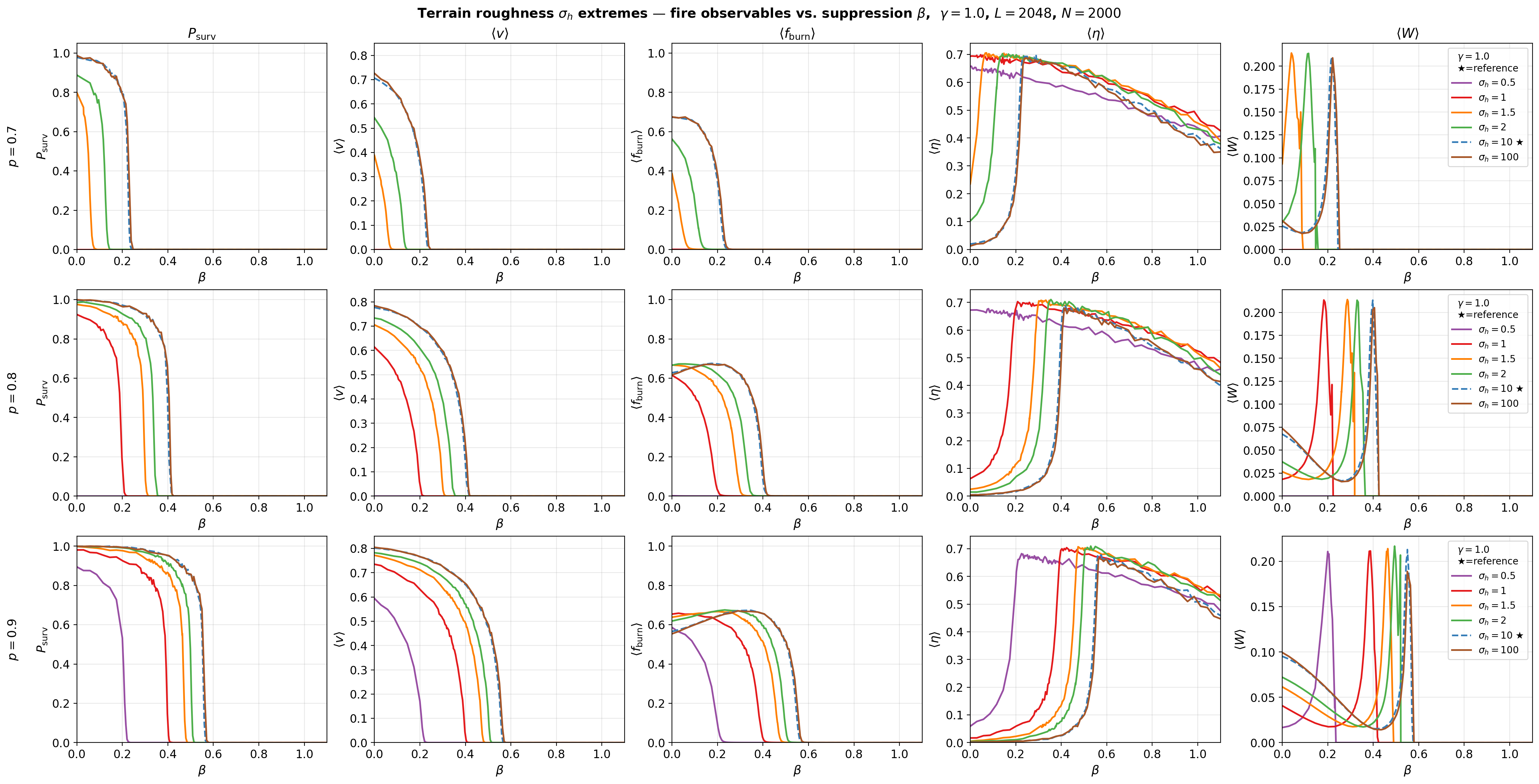}
\caption{%
  All five observables vs $\beta$ for $\sigma_h\in\{0.5,1,1.5,2,10,100\}$
  (colored lines) at $\gamma=1.0$, $L=2048$, $N=2000$;
  rows: $p=0.7$ (top), $p=0.8$ (middle), $p=0.9$ (bottom).
  Dashed ($\sigma_h=10$, $\star$): reference.
  $\betac$ decreases monotonically with decreasing $\sigma_h$ across all
  $p$; at $p=0.7$, $\sigma_h\leq 1$ gives $\betac<0$ ($\psurviv\equiv0$).
  $\avg{\eta}$ is larger for smooth terrain ($\sigma_h=10,100$) than rough
  ($\sigma_h=0.5,1$) at all densities.
}
\label{fig:phaseD}
\end{figure*}

Figure~\ref{fig:phaseD} shows all five observables for
$\sigma_h\in\{0.5,1,1.5,2,10,100\}$ at $\gamma=1.0$, for
$p\in\{0.7,0.8,0.9\}$ simultaneously.  The monotone $\betac(\sigma_h)$ is
qualitatively the same across all densities.  Shape anisotropy is consistently
larger for smooth terrain ($\sigma_h=10$, $100$) than for rough terrain
($\sigma_h=0.5$, $1$), across all $p$.

\subsection{Finite-size scaling}
\label{sec:results_fss}

\begin{figure*}[t]
\includegraphics[width=\textwidth]{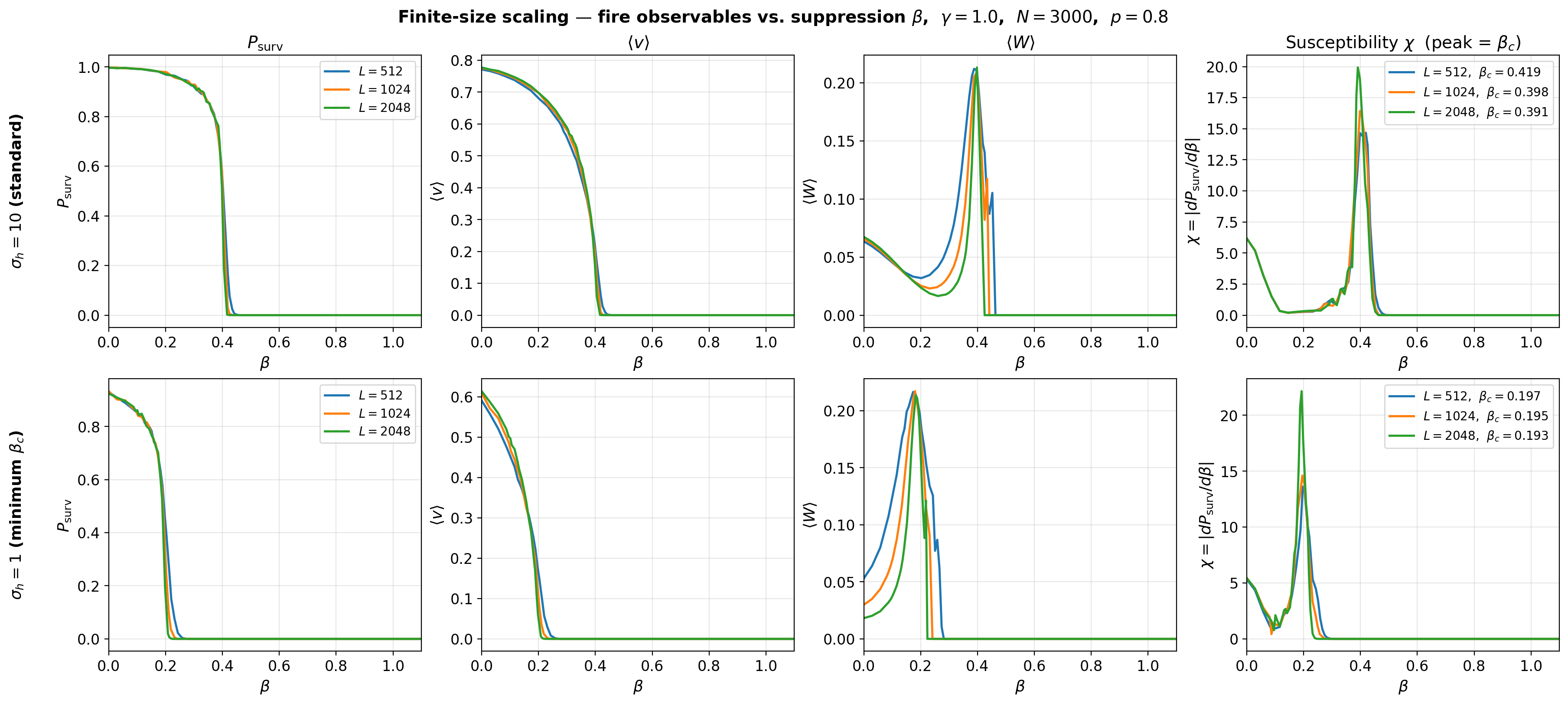}
\caption{%
  Finite-size scaling at $\gamma=1.0$, $p=0.8$, $N=3000$.
  Top row: $\sigma_h=10$; bottom row: $\sigma_h=1$.
  Columns: $\psurviv(\beta)$, $\avg{v}(\beta)$, $\avg{W}(\beta)$, and the
  susceptibility $\chi=|d\psurviv/d\beta|$ for $L\in\{512,1024,2048\}$.
  The $\betac$ values quoted in the legends of the rightmost panels are
  susceptibility-peak estimates; the $\psurviv=0.5$ crossings used in the
  text are $0.402$, $0.400$, $0.398$ ($\sigma_h=10$) and $0.195$, $0.191$,
  $0.189$ ($\sigma_h=1$).  $\betac(L)$ shifts monotonically in both cases,
  with total shift $0.004$ ($\sigma_h=10$) and $0.006$ ($\sigma_h=1$).
}
\label{fig:fss}
\end{figure*}

Figure~\ref{fig:fss} presents the FSS overview at $\gamma=1.0$, $p=0.8$,
for $L\in\{512,1024,2048\}$, $N=3000$.  Two terrain correlation lengths are
compared: $\sigma_h=10$ (smooth-terrain reference) and $\sigma_h=1$ (rough terrain).
The $\psurviv=0.5$ crossing shifts monotonically with system size for both cases,
$\betac(512)>\betac(1024)>\betac(2048)$, as expected from the finite-size
correction $\betac(L)=\betac(\infty)+aL^{-1/\nu}$.

\begin{figure*}[!t]
\includegraphics[width=\textwidth]{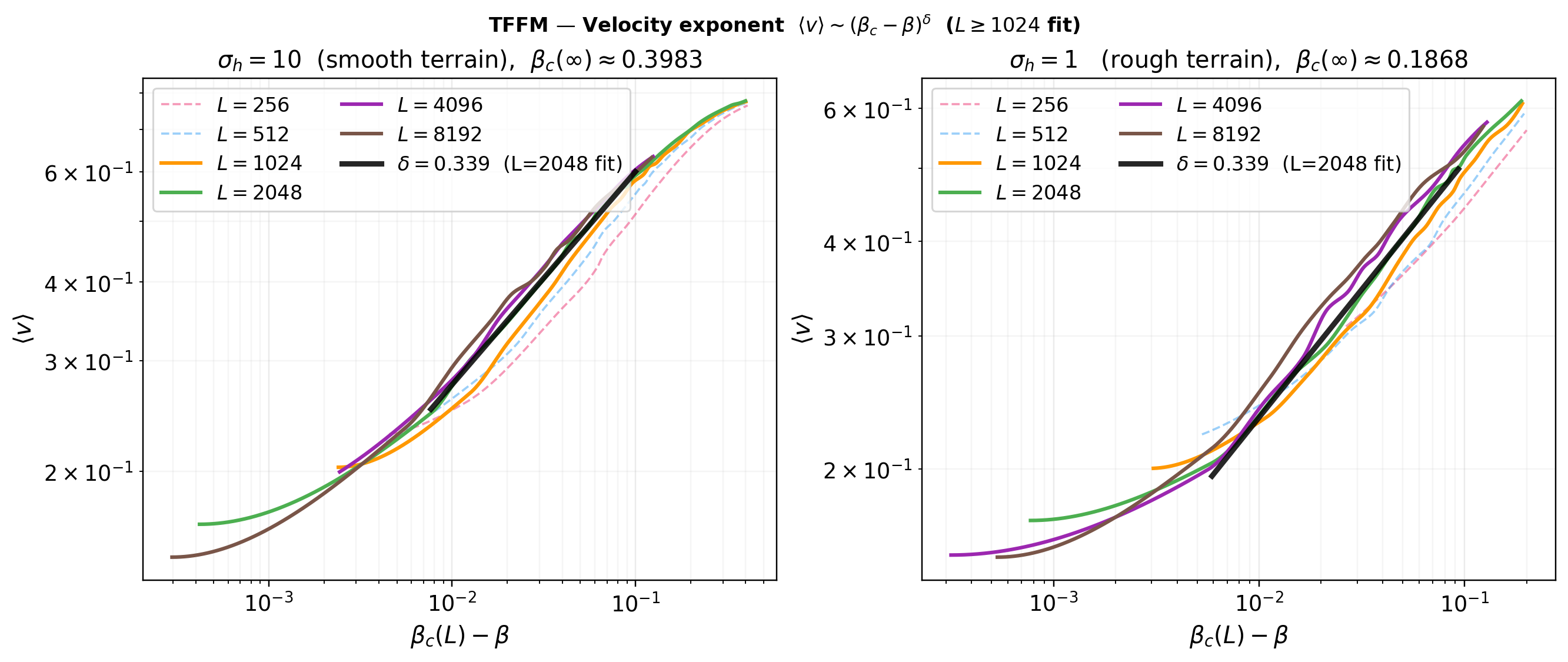}
\caption{%
  Velocity exponent at $\gamma=1.0$, $p=0.8$: $\avg{v}$ vs $\betac(L)-\beta$
  (log--log) for $L\in\{256,\dots,8192\}$ and both $\sigma_h$ ($N\geq1500$).
  $L\geq1024$ enter the power-law regime (solid); $L\leq512$ show the
  saturation plateau (dashed, excluded).  The black line is the $L=2048$ fit
  over $\betac-\beta\in[0.003,0.10]$, $\delta=0.339$; fits to all
  $L\geq1024$, both roughnesses, and both grids give
  $\delta=0.34\pm0.03$ [Eq.~(\ref{eq:delta_result})].}
\label{fig:fss_quant}
\end{figure*}

\begin{figure*}[!t]
\includegraphics[width=\textwidth]{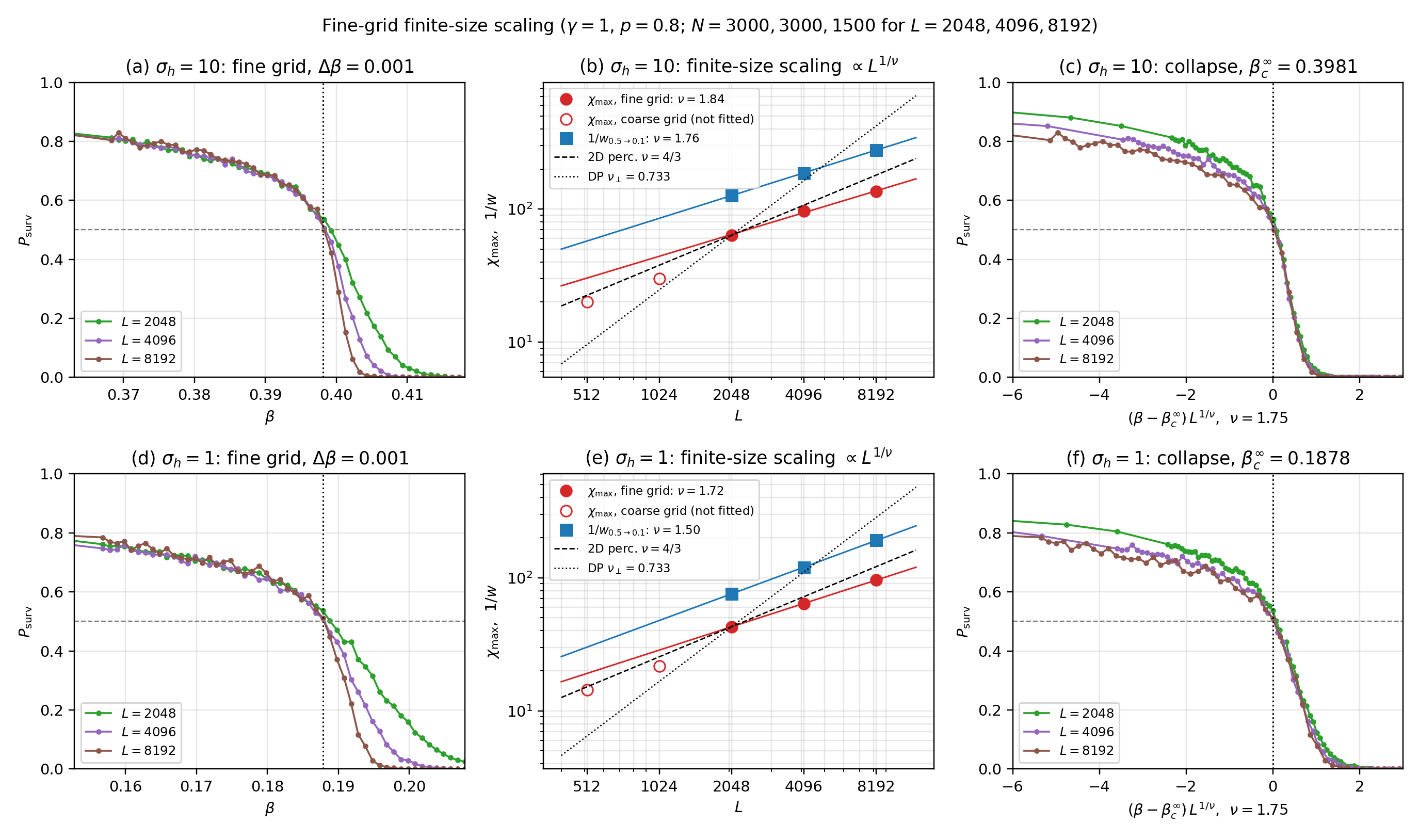}
\caption{%
  Fine-grid finite-size scaling at $\gamma=1.0$, $p=0.8$, $\Delta\beta=0.001$,
  for $L=2048$, $4096$, $8192$ ($N=3000$, $3000$, $1500$); top row
  $\sigma_h=10$, bottom row $\sigma_h=1$.  (a,d) $\psurviv(\beta)$ near
  $\betac^\infty$ (dotted line); the slow decline in the active phase is
  $L$-independent, the sharp drop from $P^*\approx0.5$ narrows with $L$.
  (b,e) Susceptibility peak $\chi_{\rm max}$ (red circles) and inverse width
  of the sharp component $1/w_{0.5\to0.1}$ (blue squares) versus $L$, with
  power-law fits giving $\nu=1.84$, $1.76$ ($\sigma_h=10$) and $1.72$,
  $1.50$ ($\sigma_h=1$); open circles are the coarse-grid $\chi_{\rm max}$
  at $L=512$, $1024$, not fitted.  Reference slopes: 2D isotropic percolation
  ($\nu=4/3$) and DP ($\nu_\perp=0.733$).  (c,f) Collapse with
  $\betac^\infty=0.3981$, $0.1878$ and $\nu=1.75$: the sharp component
  collapses; the active-phase component is not a finite-size quantity and
  does not.}
\label{fig:fss_betac}
\end{figure*}

The quantitative FSS analysis uses $L\in\{256,512,1024,2048,4096,8192\}$
for both $\sigma_h=10$ and $\sigma_h=1$ (Fig.~\ref{fig:fss_quant}).
The size-dependent critical thresholds from the
$\psurviv=0.5$ crossing are, for $\sigma_h=10$:
$\betac(256)=0.410$, $\betac(512)=0.402$,
$\betac(1024)=0.400$, $\betac(2048)=0.398$, $\betac(4096)=0.398$,
$\betac(8192)=0.398$;
and for $\sigma_h=1$:
$\betac(256)=0.200$, $\betac(512)=0.195$,
$\betac(1024)=0.191$, $\betac(2048)=0.189$, $\betac(4096)=0.188$,
$\betac(8192)=0.187$
(coarse grid, spacing $0.005$--$0.007$ near $\betac$).  On the fine grid
($\Delta\beta=0.001$, uncertainty $\pm0.0005$) the three largest sizes give
$\betac(2048)=0.3993$, $\betac(4096)=0.3984$, $\betac(8192)=0.3983$
($\sigma_h=10$) and $0.1889$, $0.1880$, $0.1880$ ($\sigma_h=1$).  Both
sequences are monotone, the total shift from $L=256$ is
$\approx0.012$--$0.013$ for both roughnesses, and the shift between
$L=4096$ and $8192$ is at the resolution limit, so $\betac^\infty$ is
resolved to $\pm0.001$.

\textit{Velocity exponent.}
The upper panels of Fig.~\ref{fig:fss_quant} show $\avg{v}\sim(\betac-\beta)^\delta$
on log--log axes.  Systems $L=256$ and $L=512$ lie in a geometric-saturation
regime: fire traverses the lattice before critical slowing-down can develop,
producing a flat velocity plateau near $\betac$ rather than a power law.
A change in slope visible at $\betac(L)-\beta\approx 0.006$--$0.010$ marks
the crossover from the critical power-law regime into this finite-size saturation;
this is a genuine physical finite-size effect rather than a numerical artifact.
Restricting fits to $L\geq1024$ over the range
$\betac(L)-\beta\in[0.003,0.10]$ (about $1.5$ decades) yields
\begin{equation}
  \delta = 0.34\pm 0.03,
  \label{eq:delta_result}
\end{equation}
where the uncertainty combines the statistical fit error with the spread
between system sizes, between the two terrain roughnesses, and between fit
windows.  On the coarse grid the single-size fits give $\delta=0.34$, $0.34$,
$0.32$ for $L=2048$, $4096$, $8192$ at $\sigma_h=10$ and $0.34$, $0.35$,
$0.33$ at $\sigma_h=1$.  On the fine grid, which samples the saturation
plateau densely, the lower end of the window must be moved to
$\betac(L)-\beta=0.01$ to exclude it; the fits then give $0.38$, $0.35$,
$0.32$ ($\sigma_h=10$) and $0.35$, $0.36$, $0.35$ ($\sigma_h=1$).  Using
$\betac^\infty$ instead of $\betac(L)$ in the abscissa changes these by at
most $0.02$.  We therefore report $\delta\approx0.34$ as the velocity
exponent of the \TFFM\ critical point, robust against terrain roughness.  We
note that $\avg{v}$ is defined on surviving runs only and measured at
boundary-crossing time (Sec.~\ref{sec:obs}), so $\delta$ characterizes the
asymptotic front velocity in the active phase rather than the transient
spreading from a seed.

\textit{Critical threshold in the thermodynamic limit.}
The pairwise-crossing method (Sec.~\ref{sec:betac_method}) applied to
consecutive size pairs gives crossings that drift monotonically upward and
converge: for $\sigma_h=10$, $\beta_\times=0.3926$, $0.3969$, $0.3983$ for
the pairs $(1024,2048)$, $(2048,4096)$, $(4096,8192)$, with $\psurviv$ at
the crossing decreasing from $0.64$ to $0.55$ and $0.47$; for $\sigma_h=1$,
$\beta_\times=0.1795$, $0.1812$, $0.1868$ with $\psurviv$ at the crossing
$0.63$, $0.61$, $0.52$.  The crossing value thus approaches $P^*\approx0.5$,
and the largest pair gives
\begin{equation}
\begin{gathered}
  \betac^\infty = 0.398\pm0.001 \quad (\sigma_h=10), \\
  \betac^\infty = 0.188\pm0.001 \quad (\sigma_h=1);
\end{gathered}
  \label{eq:betac_inf}
\end{equation}
on the fine grid the $(4096,8192)$ crossing lies at $0.3981$ and $0.1878$
with $\psurviv=0.51$, and coincides with the single-size $\psurviv=0.5$
crossings at $L\geq4096$.  A two-parameter fit of the fine-grid
$\betac(L)=\betac^\infty+aL^{-1/\nu}$ with $\nu=1.75$ (below) extrapolates
to $0.3974$ and $0.1871$, one grid spacing lower; we quote the crossing
value.

\textit{Fine-grid finite-size scaling and the correlation-length exponent.}
Figure~\ref{fig:fss_betac} shows the finite-size scaling analysis on the
fine $\beta$ grid at $L=2048$, $4096$, and $8192$.  Panels (a,d) display the
transition profiles.  They have two components.  Across the active phase
$\psurviv$ declines smoothly and slowly with $\beta$ (from $0.83$ to
$0.55$ over the range shown for $\sigma_h=10$), and this decline is
\emph{independent of $L$}: it is the intrinsic survival probability of the
active phase, not a finite-size effect.  Superimposed on it is the sharp drop
from $P^*\approx0.5$ to zero, whose width shrinks with $L$ and which carries
the finite-size scaling.  At $\betac^\infty$ itself $\psurviv=0.54$,
$0.51$, $0.51$ ($\sigma_h=10$) and $0.53$, $0.51$, $0.51$ ($\sigma_h=1$)
for the three sizes.

Panels (b,e) show two estimators of $\nu$ built from the sharp component.
The raw susceptibility peak $\chi_{\rm max}=\max|d\psurviv/d\beta|$ grows
as $64\to96\to135$ ($\sigma_h=10$) and $43\to64\to96$ ($\sigma_h=1$), i.e.,
by a factor $1.4$--$1.5$ per doubling, giving $\nu=1.84$ and $1.72$ from
power-law fits ($1.67$, $2.04$ and $1.73$, $1.71$ from consecutive pairs).  The
width of the sharp component, defined as the interval between
$\psurviv=0.5$ and $0.1$, decreases as $0.0079\to0.0054\to0.0036$
($\sigma_h=10$) and $0.0132\to0.0084\to0.0052$ ($\sigma_h=1$), giving
$\nu=1.76$ and $1.50$.  The two coarse-grid sizes measured on a common grid,
$L=512$ and $1024$, give $\nu=1.7$ for both roughnesses and fall on the
extrapolated fine-grid line (open symbols).  Combining the estimators and
roughnesses,
\begin{equation}
  \nu = 1.8\pm0.2 ,
  \label{eq:nu_result}
\end{equation}
where the uncertainty spans the values obtained from the two estimators and
the two terrain roughnesses.  This value is clearly distinct from two-dimensional
isotropic percolation ($\nu=4/3$) and from the spatial exponent of
$(2{+}1)$-dimensional directed percolation ($\nu_\perp=0.733$), both drawn as
reference slopes in the figure.

Panels (c,f) show the data collapse of the fine-grid profiles with
$\betac^\infty$ from Eq.~(\ref{eq:betac_inf}) and $\nu=1.75$.  The sharp
component collapses onto a single curve for all three sizes, while the slow
active-phase component does not and should not: it is an $L$-independent
function of $\beta$, not of $(\beta-\betac)L^{1/\nu}$.  The collapse thus
confirms the picture of a survival probability that jumps from
$P^*\approx0.5$ to zero at $\betac$, with finite-size rounding governed by a
conventional correlation-length exponent~\cite{Henkel2008,Li2023}.

A methodological remark is in order, because the fine-grid result reverses
the conclusion one would draw from the coarse grid alone.  On the coarse grid
(spacing $0.005$--$0.007$, i.e., only three to six points across the sharp
component at $L\geq4096$) the susceptibility peak is resolution limited:
$\chi_{\rm max}$ saturates at $\approx58$ ($\sigma_h=10$) and $\approx50$
($\sigma_h=1$) for $L\geq4096$, the fitted $\nu$ depends on any smoothing
applied before differentiation ($2.2$--$2.5$ raw, $3.9$ after smoothing over
$1.5$ grid spacings, $5$--$6$ after $3$), and consecutive-pair estimates
mixing the two grids scatter between $1.3$ and $42$.  None of this is a
property of the model.  With a grid several times finer than the transition
width at every size, the estimators agree with each other and across
roughnesses.  Quenched, spatially correlated disorder is known to generate
slow crossover corrections to scaling~\cite{Weinrib1983,Vojta2006,PastorSatorras2000},
and we cannot exclude a further drift of $\nu$ beyond $L=8192$; but the
three sizes measured on the fine grid span two octaves with no systematic
trend.

\begin{figure*}[!t]
\includegraphics[width=\textwidth]{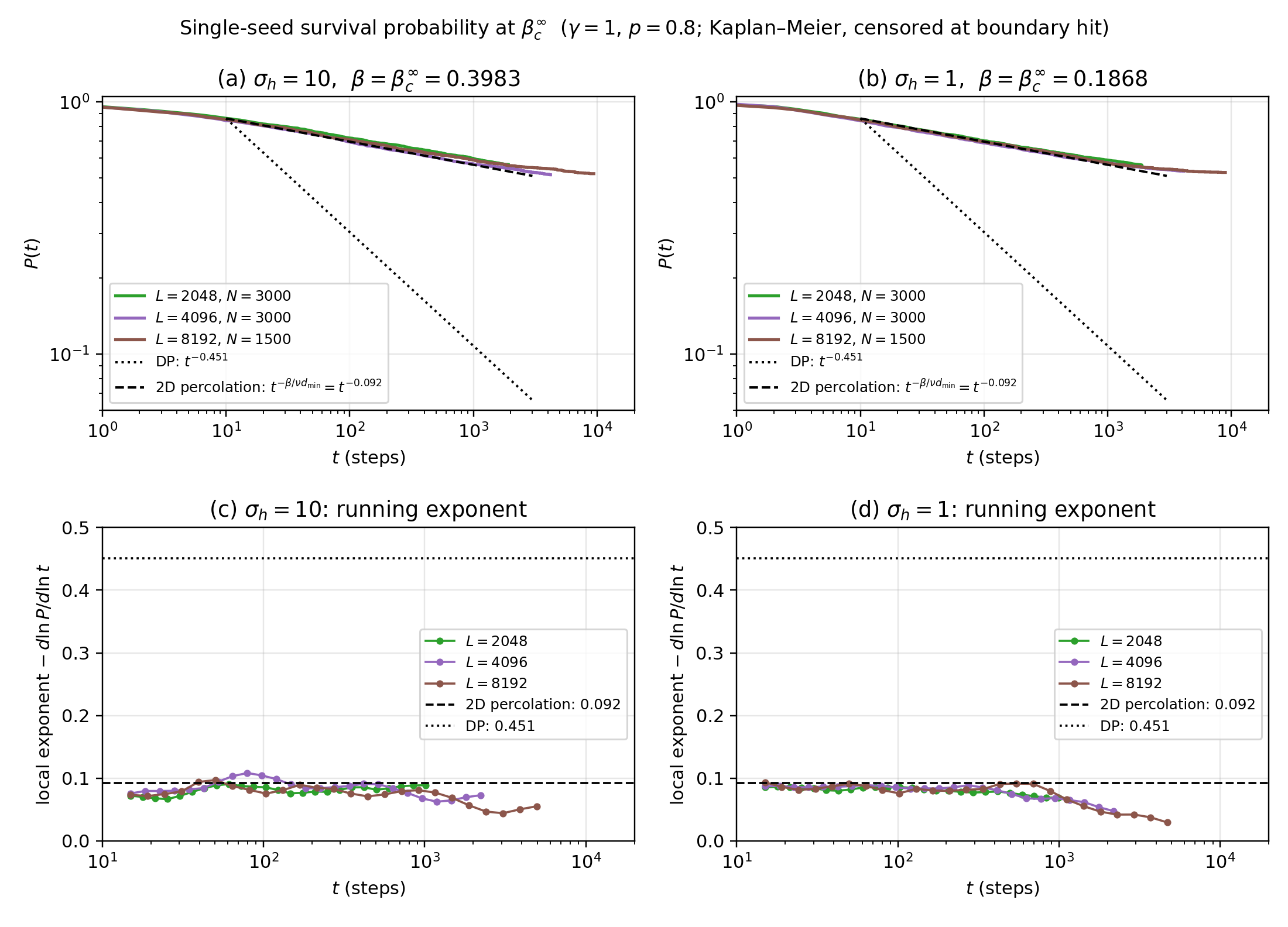}
\caption{%
  Time-dependent survival probability of a single-seed fire at
  $\beta=\betac^\infty$ ($\gamma=1.0$, $p=0.8$).  (a,b) $P(t)$ for
  $L\in\{2048,4096,8192\}$ with $N=3000$, $3000$, $1500$ realizations, for
  $\sigma_h=10$ and $\sigma_h=1$; runs reaching the absorbing boundary are
  right-censored, and each curve is shown up to the earliest boundary hit.
  Reference slopes: directed percolation ($t^{-0.451}$) and two-dimensional
  isotropic percolation ($t^{-0.092}$).  (c,d) Running exponent
  $-\,d\ln P/d\ln t$ from sliding log-window fits; the dashed line is the
  isotropic-percolation value.  The exponent decreases steadily with $t$ and
  the curves for different $L$ coincide, indicating a slow approach to a
  finite $P_\infty$ rather than a power-law decay to zero.}
\label{fig:pt}
\end{figure*}

\textit{Survival probability at criticality.}
Figure~\ref{fig:pt} shows $P(t)$ at $\beta=\betac^\infty$ for
$L\in\{2048,4096,8192\}$.  Three features stand out.  First, the curves for
different $L$ coincide within statistical error over their common range
(e.g., $P(1000)=0.595$, $0.568$, $0.589$ for $\sigma_h=10$), so the decay is
a property of the infinite system, not of the boundary.  Second, the decay is
extremely slow: $P(t)$ falls only from $0.85$ at $t=10$ to $0.52$ at
$t\approx8000$.  A running exponent
$-\,d\ln P/d\ln t$ (panels c,d) is $\approx0.09$ for
$t\in[30,300]$ and then decreases steadily to $0.05$ ($\sigma_h=10$) and
$0.03$ ($\sigma_h=1$) over $t\in[3000,8000]$; a pure power law does not fit
(rms residual $0.007$--$0.009$), whereas $P(t)=P_\infty+At^{-a}$ fits with
rms $0.003$--$0.005$ and $P_\infty=0.35\pm0.03$, $a\approx0.17$--$0.19$ for
both roughnesses at $L=8192$.  Third, the extinction-time distribution is
strongly bimodal: of the fires that die, $60$--$65\%$ do so within the first
$100$ steps, and the median extinction time ($\approx30$--$50$ steps) is
independent of $L$.

The directed-percolation value $\delta_{\rm DP}=0.451$ is excluded by a
factor of five at every $t$ and every $L$.  The isotropic-percolation value
$0.092$ describes the intermediate window $t\in[30,300]$ but not the
continued flattening beyond it.  The data are instead most naturally read as
a survival probability that tends to a finite limit $P_\infty$ in the
critical state: a fire that escapes the local trap structure around the
ignition point within the first $\mathcal{O}(10^2)$ steps is, with high
probability, never extinguished.  This is the dynamical counterpart of the
static observation that $\psurviv$ at the transition takes the
$L$-independent value $P^*\approx0.5$ before dropping to zero
(Fig.~\ref{fig:fss_betac}), and of the two-component profile of
Fig.~\ref{fig:fss_betac}(a,d).  We cannot exclude, with $t\lesssim10^4$, that
$P(t)$ eventually vanishes as an ultra-slow power law; a definitive test
requires $L\gtrsim3\times10^4$.

\subsection{Wind-driven fire: phase boundary, anisotropy, and drift}
\label{sec:results_wind}

\begin{figure*}[t]
\includegraphics[width=\textwidth]{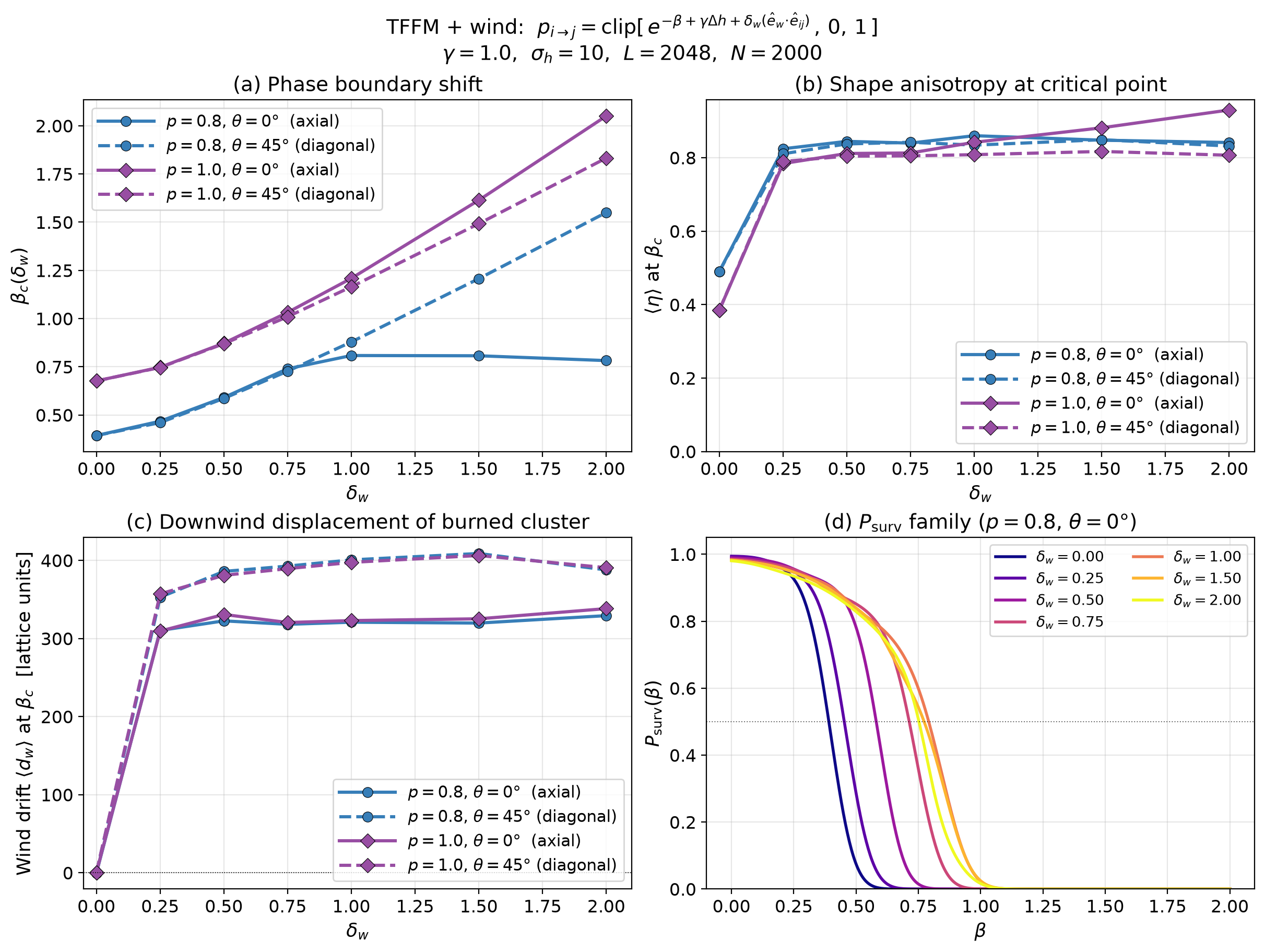}
\caption{%
  Wind coupling in the \TFFM\ ($\gamma=1.0$, $\sigma_h=10$, $L=2048$,
  $N=2000$). Lines: $p\in\{0.8,1.0\}$ (blue, purple);
  $\theta\in\{0^{\circ},45^{\circ}\}$ (solid, dashed).
  (a)~$\betac(\dw)$: monotonically increasing; axial wind at $p=0.8$ saturates
  at $\approx 0.81$ once downwind bonds clip to unity; diagonal wind and
  $p=1.0$ reach $\betac\approx 1.83$ at $\dw=2.0$.
  (b)~$\avg{\eta}$ at $\betac$: rises from the terrain-only baseline
  ($\approx 0.4$--$0.5$) to $\approx 0.85$, saturating at $\dw\gtrsim 0.75$.
  (c)~Drift $\avg{d_w}$ at $\betac$: sharp onset at $\dw=0.25$ to
  $\approx 310$ (axial) and $\approx 355$ (diagonal) lattice units, then a
  plateau at $\approx 320$--$330$ (axial) and $\approx 385$--$405$
  (diagonal).
  (d)~$\psurviv(\beta)$ for $p=0.8$, $\theta=0^{\circ}$,
  $\dw\in\{0,\ldots,2.0\}$: rightward shift; $\dw=1.5$ and $2.0$ coincide.
}
\label{fig:wind_summary}
\end{figure*}

Wind coupling introduces three qualitatively distinct effects into the
\TFFM: an amplification of the fire risk threshold $\betac$ by a factor of
up to $\approx4$, rapid saturation of fire-scar shape anisotropy, and a
sharp onset of downwind drift at even the weakest coupling.  Figure~\ref{fig:wind_summary}
shows all four observables at $\gamma=1.0$, $\sigma_h=10$, $L=2048$, $N=2000$.

\textit{Phase boundary.}
Panel~(a) shows $\betac(\dw)$ for both wind directions and both tree densities.
$\betac$ increases monotonically with $\dw$ across all four combinations:
wind-driven fire requires greater suppression to halt, raising the fire risk
threshold systematically above the terrain-only value.

Two qualitatively distinct regimes appear.  For \textit{axial} wind
($\theta=0^{\circ}$, east) at $p=0.8$, $\betac$ rises from $0.40$ at $\dw=0$
to $\approx 0.81$ at $\dw=1.0$, beyond which it saturates.  This saturation
reflects a clipping effect: once essentially all eastward bonds are clipped to
probability~1, additional wind no longer extends the downwind fire reach.  The
limiting factor then becomes forest connectivity ($p=0.8$, 20\% empty sites).
For \textit{diagonal} wind ($\theta=45^{\circ}$), the same mechanism applies
to two bond directions simultaneously but each receives only $\dw/\sqrt{2}$
bias, so clipping requires larger $\dw$ and $\betac$ grows continuously,
reaching $\approx 1.56$ at $\dw=2.0$.

At $p=1.0$ (fully connected forest), neither wind direction saturates within
the studied range; diagonal wind drives $\betac$ to $\approx 1.83$ at
$\dw=2.0$---$2.7$ times the no-wind baseline of $0.68$.  Panel~(d)
(Fig.~\ref{fig:wind_summary}) shows the $\psurviv(\beta)$ family for
$p=0.8$, $\theta=0^{\circ}$, where curves for $\dw=1.5$ and $\dw=2.0$
nearly coincide at $\betac\approx 0.80$, confirming phase-boundary saturation.

\textit{Shape anisotropy.}
Panel~(b) shows $\avg{\eta}$ at $\betac$ as a function of $\dw$.
At $\dw=0$, the fire scars already carry terrain-induced anisotropy:
$\avg{\eta}\approx 0.50$ for $p=0.8$ and $\approx 0.39$ for $p=1.0$.
Wind immediately enhances this anisotropy, which saturates at
$\avg{\eta}\approx 0.85$ already at $\dw\gtrsim 0.75$, independently of
wind direction and tree density.  Once downwind bonds enter the clipping regime,
the shape of the fire scar is determined by the geometry of the burning cluster
rather than the wind strength.  For an elliptical cluster,
$\avg{\eta}\approx 0.85$ implies a major-to-minor axis ratio
$a/b=(1-\eta)^{-1/2}\approx 2.6$, placing wind-driven \TFFM\ scars in the
strongly elongated regime
documented in field studies of plantation fires under sustained
wind~\cite{DirectionalFire2025}.

\textit{Wind-direction drift.}
Panel~(c) shows the most distinctive signature of wind coupling: the downwind
center-of-mass displacement $\avg{d_w}$ at $\betac$.  At $\dw=0$,
$\avg{d_w}=0$ by symmetry.  At $\dw=0.25$---the smallest nonzero wind
coupling---the drift jumps immediately to $\approx 310$ lattice units for
axial wind and $\approx 355$ for diagonal wind, essentially independently of
$p$ (in a system of linear size $L=2048$, so $L/2=1024$ sites from center to
boundary).  This sharp onset is followed by a plateau: for all $\dw\geq 0.5$
the drift remains nearly constant at $\approx 320$--$330$ lattice units
(axial) and $\approx 385$--$405$ lattice units (diagonal).  The larger
diagonal values reflect the longer path, by a factor $\sqrt{2}$, from the
center to the corner of the square domain.  The onset-then-plateau pattern
reflects a threshold phenomenon: even a weak wind immediately selects a
downwind fire pathway that displaces the scar center of mass by
$\sim\!30$--$40\%$ of the half-system size.

Together, the risk-threshold amplification, the shape saturation, and the
drift onset provide complementary diagnostics for wind coupling that span
the full parameter range: $\betac$ is most sensitive to strong wind,
$\avg{\eta}$ saturates early and signals the clipping regime, and $\avg{d_w}$
is most sensitive to weak onset.

\subsection{Fire-scar morphology: terrain structure, wind coupling, and the paradox}
\label{sec:results_morph}

The statistical observables of the preceding subsections are complemented
by direct visualization of fire scars at the mesoscale ($L=200$), which make
the interplay between terrain structure and wind coupling immediately apparent.
In each panel below, yellow isochrones mark equal-time propagation fronts,
the orange curve delimits the scar boundary, and the cyan star marks the
ignition site; interior micro-patches at $p=0.8$ reflect the site-dilution
effect of unoccupied lattice sites.

\begin{figure*}[t]
\includegraphics[width=\textwidth]{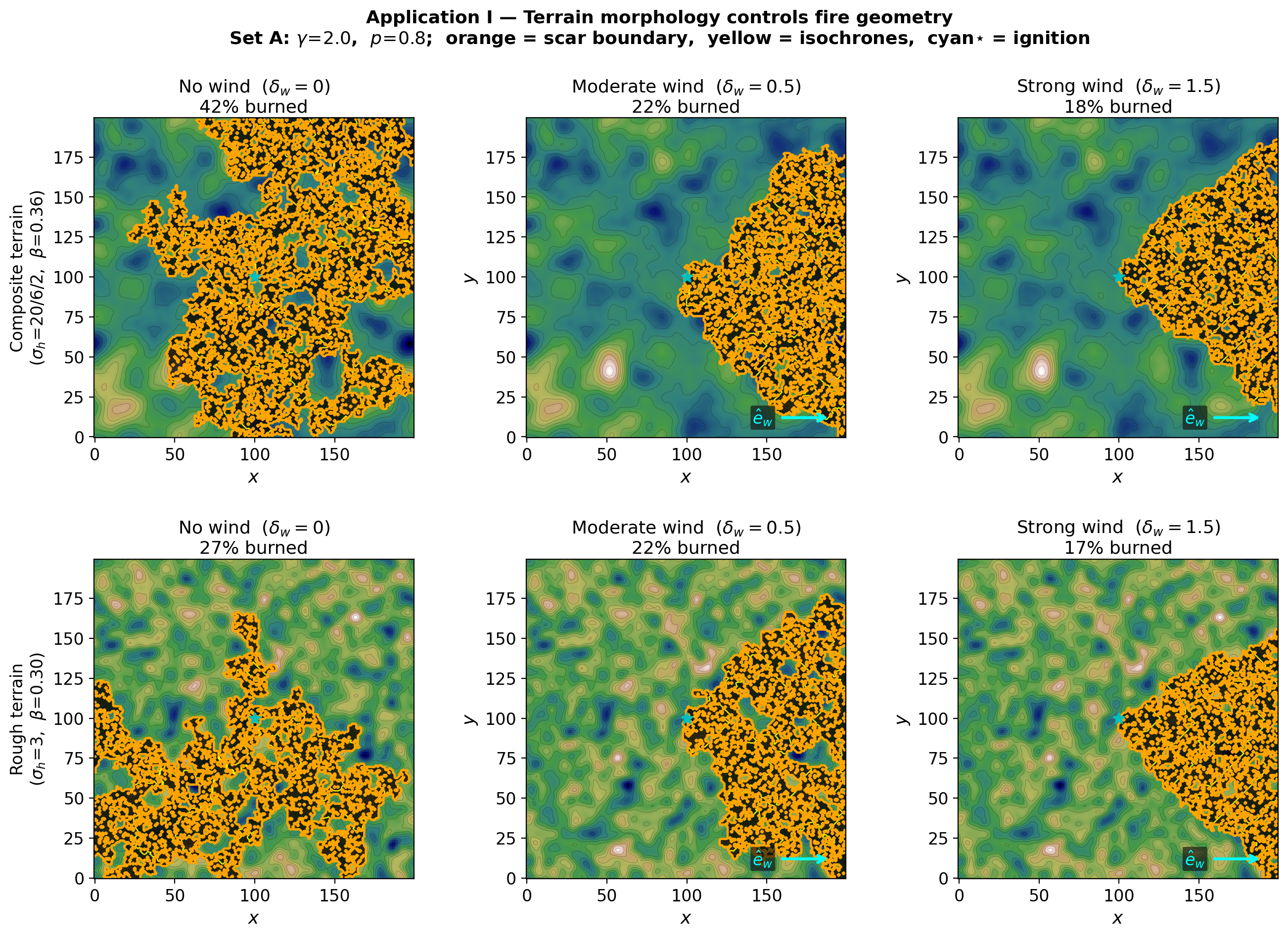}
\caption{%
  \textbf{Terrain morphology controls fire-scar geometry (Set~A, $\gamma=2.0$,
  $p=0.8$).}  Single near-critical realization at $L=200$; orange = scar
  boundary, yellow = isochrones, cyan $\star$ = ignition, cyan arrow = east wind.
  Top: composite terrain ($\sigma_h=20/6/2$, $\beta=0.36$);
  bottom: rough terrain ($\sigma_h=3$, $\beta=0.30$).
  Columns: $\dw=0$, $0.5$, $1.5$.
  Composite terrain produces elongated, ridge-following scars that stretch
  progressively downwind; rough terrain fragments the boundary and multiplies
  interior micro-patches.
}
\label{fig:app_morph}
\end{figure*}

Figure~\ref{fig:app_morph} contrasts two terrain archetypes across three east-wind
intensities for Set~A ($\gamma=2.0$, $p=0.8$), with each panel run near its own
$\betac$.  The top row uses composite multi-scale terrain ($\sigma_h=20/6/2$);
the bottom row uses single-scale rough terrain ($\sigma_h=3$).  On composite
terrain, system-spanning ridge-and-valley structure produces elongated,
directionally coherent scars even without wind ($\dw=0$, left column); as east
wind increases, the scar stretches further downwind, isochrones compress on the
lee side, and the burned fraction decreases slightly as the front exits the
eastern boundary before filling the domain.  Rough terrain, by contrast,
fragments the scar perimeter and multiplies unburned micro-patches: short-range
height fluctuations at small $\sigma_h$ lack the coherent directional channels
that sustain elongated propagation.  Reading left to right along either row
quantifies the wind contribution at fixed terrain; reading top to bottom within
any column isolates the terrain contribution at fixed wind.

\begin{figure*}[t]
\includegraphics[width=\textwidth]{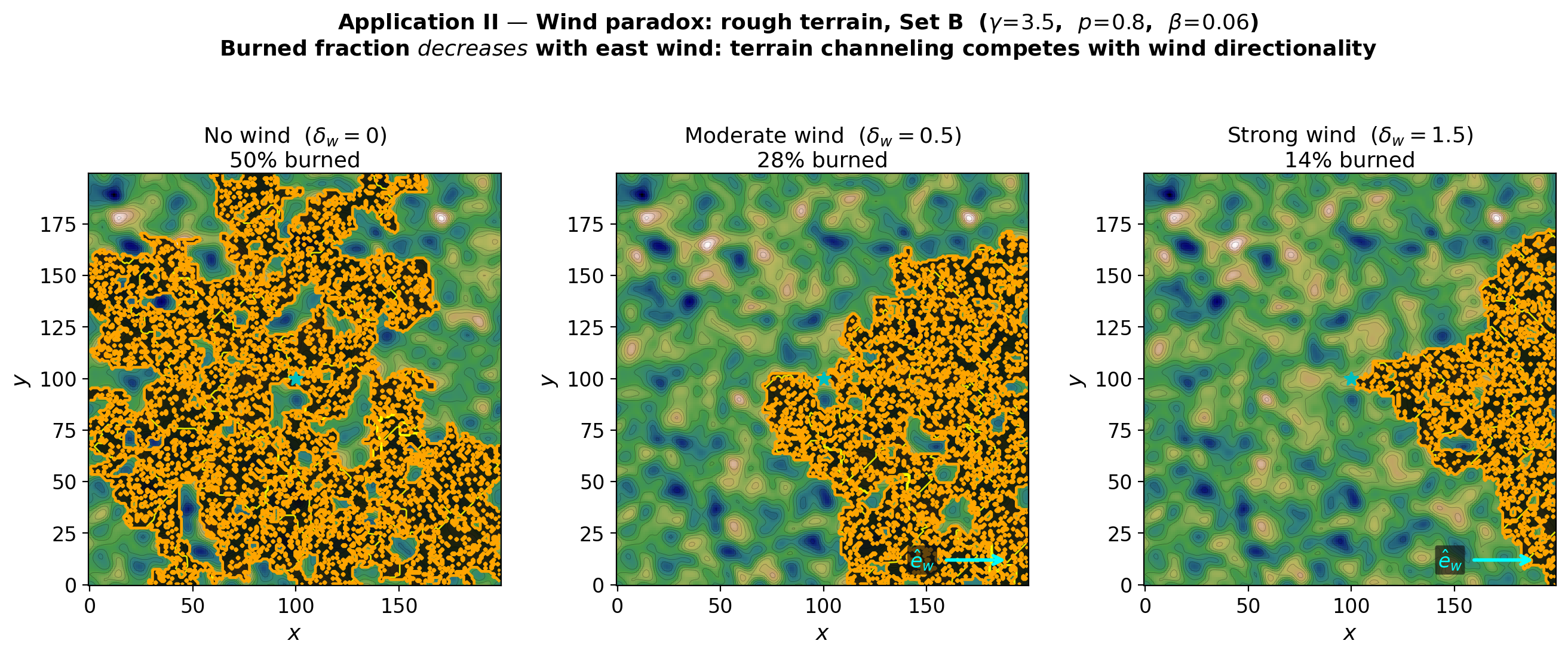}
\caption{%
  \textbf{Wind paradox: terrain-wind competition (Set~B, $\gamma=3.5$,
  $p=0.8$, $\beta=0.06$, $\sigma_h=3$).}
  Same terrain seed for three east-wind intensities.
  Burned fraction at boundary crossing \emph{decreases} monotonically:
  $\approx 50\%$ ($\dw=0$) $\to$ $28\%$ ($\dw=0.5$) $\to$ $14\%$ ($\dw=1.5$).
  East wind funnels the front into a narrow downwind corridor that exits the
  boundary before the terrain-driven radial expansion fills the domain,
  reducing total burned area despite stronger wind.
  This inversion has no counterpart in the \WFFM\ or in operational models
  treating terrain and wind additively.
}
\label{fig:app_paradox}
\end{figure*}

Figure~\ref{fig:app_paradox} isolates the terrain-wind paradox on rough terrain
for Set~B ($\gamma=3.5$, $p=0.8$, $\beta=0.06$, $\sigma_h=3$), using the same
random seed across all three wind intensities so that the terrain realization is
identical and only the wind coupling varies.  The burned fraction at
boundary crossing decreases monotonically: $\approx 50\%$ at $\dw=0$,
$\approx 28\%$ at $\dw=0.5$, and $\approx 14\%$ at $\dw=1.5$---the
opposite of the naive expectation that stronger wind enlarges fire.  The isochrones reveal the
mechanism: without wind, the fire front expands radially via correlated terrain
pathways in all directions; as east wind increases, the front is progressively
funneled into a narrow downwind corridor that exits the eastern boundary before
the terrain-driven radial expansion can fill the domain, yielding a more
elongated but smaller scar.  This inversion is absent from the isotropic
\WFFM\ and from operational models that treat terrain and wind as purely
additive accelerators~\cite{Finney1998,Cruz2017}.

\section{Discussion}
\label{sec:discussion}

\subsection{Comparison with the WFFM}
\label{sec:disc_wffm}

The \WFFM~\cite{wffm2025} and the \TFFM\ share the same lattice geometry,
state space, synchronous update rule, and velocity observable, and both exhibit
a phase transition between active and inactive regimes controlled by $\beta$.
Their bond-disorder structures, however, differ fundamentally.

In the \WFFM, disorder is \emph{symmetric} and \emph{spatially uncorrelated}:
the active phase is sustained by rare high-weight bonds that form percolating
clusters, pushing $\betac$ to large values ($\approx 1.01$ at $p=0.8$~\cite{wffm2025}).
In the \TFFM, the disorder is \emph{asymmetric} and \emph{spatially correlated}:
fire propagates preferentially along coherent uphill pathways and is held
back on descending flanks, so that ridgelines act as firebreaks and the
active phase is sustained only where the network of favorable bonds remains
connected.  The terrain correlation length $\sigma_h$ has no analog in the
\WFFM\ and controls the spatial extent of these pathways and barriers.  A third difference is observable:
the broken rotational symmetry of the \TFFM\ produces anisotropic fire shapes,
quantified by $\avg{\eta}$, which the \WFFM\ lacks entirely.

\subsection{Physical origin of the monotone $\betac(\sigma_h)$ dependence}
\label{sec:disc_minimum}

For large $\sigma_h$, the terrain is smooth: adjacent height differences
$h_j-h_i$ are $\mathcal{O}(1/\sigma_h)$, so $e^{\gamma(h_j-h_i)}\approx 1$ and
the terrain factor contributes negligibly.  Fire spreads essentially at the
bare rate $e^{-\beta}$, and $\betac$ approaches the $\gamma=0$ value.

For $\sigma_h=0$ (white noise), adjacent site heights are uncorrelated and the
RMS height difference between neighbors is maximal, $\sqrt{2}$ in units of
the field amplitude.  Half of all bonds are then strongly suppressed
($\Delta h<0$) while the other half are enhanced only up to the clipping
value $p_{i\to j}=1$.  The asymmetry is decisive: every site that is a local
maximum of $h$ has all four outgoing bonds penalized, and for white noise
one site in five is a local maximum.  Fire that reaches such a site is
trapped, so the effective connectivity of the spreading network is strongly
reduced and $\betac$ reaches its minimum.  A mean-field estimate illustrates
the magnitude: at $\beta=0$, $\gamma=1$, the average outgoing probability is
$\avg{\operatorname{clip}[e^{\Delta h}]}\approx0.71$ for $\Delta h\sim
\mathcal{N}(0,2)$, i.e., white-noise terrain alone is as suppressive as a
flat-terrain value $\beta\approx0.34$, comparable to the observed
$\betac(0)=0.287$ at $p=1$.

For intermediate $\sigma_h$, local maxima become rarer (their density scales
as $\sigma_h^{-2}$) and the penalized descending flanks organize into
extended ridgelines.  Fire can now climb coherently along many-site uphill
pathways before encountering a barrier, and as $\sigma_h$ increases these
pathways lengthen while the bond-scale penalty weakens, monotonically raising
$\betac$ toward the smooth-terrain plateau.  For sufficiently small
$\sigma_h$ and $p$ (Table~\ref{tab:betac_sigmah}) the combination of site
dilution and slope trapping disconnects the network entirely and $\betac$
falls below zero: no fire percolates even at $\beta=0$.

\subsection{Universality and critical behavior}
\label{sec:disc_universality}

The question of universality class is central for any percolation model.
The \WFFM~\cite{wffm2025} reports a transition consistent with directed
percolation (DP) universality in the velocity exponent, with finite-size
corrections characterized by the exponent $0.83$.

The \TFFM\ introduces \emph{locally directed} disorder: the terrain creates
preferred spreading directions that vary from site to site over the scale
$\sigma_h$---unlike standard DP models, which have a single global preferred
direction, and unlike isotropic bond percolation, in which no bond is
directed.  At long wavelengths the statistical symmetry of the Gaussian
terrain under $h\to -h$ is restored, so the long-wavelength theory need not
be DP; but because the local bias is quenched and asymmetric, it need not be
isotropic percolation either.

We begin with the velocity exponent,
$\delta=0.34\pm0.03$, stable for $L\geq1024$ and identical within errors for
$\sigma_h=10$ and $\sigma_h=1$.  To place it in context we compare with the
exponent governing the asymptotic front velocity in the two reference
classes, since $\avg{v}$ is defined at boundary crossing and therefore probes
the linear-growth regime of the active phase.  In DP the radius of a
surviving cluster grows as $R\sim\xi_\perp\,t/\xi_\parallel$, giving
$v\sim(\beta_c-\beta)^{\nu_\parallel-\nu_\perp}$ with
$\nu_\parallel-\nu_\perp\approx1.295-0.733=0.56$ in $(2{+}1)$
dimensions~\cite{Hinrichsen2000,Odor2004,Hof2023}.  In isotropic
percolation with synchronous (burning-algorithm) dynamics the front advances
along chemical paths, so $v\sim\xi^{\,1-d_{\min}}\sim(p-p_c)^{\nu(d_{\min}-1)}$
with $d_{\min}\approx1.13$ and $\nu=4/3$ in two
dimensions~\cite{Stauffer1994,Grassberger1992}, i.e., an exponent
$\approx0.18$.  The \TFFM\ value lies between these two, roughly a factor of
two away from each, and the difference is far larger than the quoted
uncertainty.  The frequently quoted DP density-decay exponent
$\delta_{\rm DP}\approx0.45$ refers to a different observable and is not
the appropriate comparison for a front velocity.

The time-dependent survival probability (Sec.~\ref{sec:results_fss},
Fig.~\ref{fig:pt}) sharpens this comparison considerably.  Directed
percolation predicts $P(t)\sim t^{-0.451}$ at criticality; the \TFFM\ data
decay five to ten times more slowly at every time and size studied, so DP is
excluded by two independent observables ($\delta$ and $P(t)$) without
reference to $\nu$.  Isotropic percolation predicts $P(t)\sim t^{-0.092}$,
which matches the running exponent over one decade but not its continued
decrease to $0.03$--$0.05$ at $t\sim10^4$; and it predicts a front-velocity
exponent of $0.18$, half the measured $\delta$.  The simplest reading of all
the evidence---the $L$-independent $P^*\approx0.5$ at which $\psurviv$
jumps to zero, the two-component transition profile of
Fig.~\ref{fig:fss_betac}(a,d), the bimodal extinction times, and the flattening
of $P(t)$---is that the survival probability of the \TFFM\ is
\emph{discontinuous} at $\betac$: at criticality a finite fraction
$P_\infty\approx0.3$--$0.5$ of single-seed fires never dies.  Physically,
the fate of a fire is decided within the first $\mathcal{O}(10^2)$ steps by
whether it escapes the trap structure (local elevation maxima and downhill
bonds) surrounding the ignition site; once it has connected to the
system-spanning network of favorable bonds, the residual suppression cannot
stop it.  Neither DP nor ordinary percolation has this property, both having
$P(t)\to0$ at criticality, whereas a discontinuous order parameter with a
diverging length scale is known from compact directed percolation and from
some models with quenched disorder~\cite{Hinrichsen2000,Vojta2006}.  We
state this as the most plausible interpretation rather than an established
result, because with $t\lesssim10^4$ an ultra-slow power-law decay cannot
be ruled out, and because $\delta$ itself may be an effective exponent of
a slow crossover, as is common in quenched-disorder
models~\cite{Vojta2006,PastorSatorras2000}.

The correlation-length exponent completes the picture.  With a $\beta$ grid
several times finer than the transition width at $L=2048$--$8192$, the
susceptibility peak and the width of the sharp component of $\psurviv$ give
consistently $\nu=1.8\pm0.2$ for both terrain roughnesses
(Sec.~\ref{sec:results_fss}).  This is neither the isotropic-percolation
value $4/3$ nor the DP value $\nu_\perp=0.733$, and it is also different from
compact directed percolation ($\nu_\perp=1$, $\nu_\parallel=2$), the
best-known example of an absorbing-state transition with a discontinuous
survival probability~\cite{Hinrichsen2000}.  The \TFFM\ critical point is
therefore characterized by three quantities---$\nu\approx1.8$,
$\delta\approx0.34$, and a survival probability that appears to remain finite
at criticality---none of which matches the reference classes.  We refrain
from declaring a new universality class on the strength of two octaves in
$L$, since quenched correlated disorder can produce effective exponents that
drift slowly~\cite{Weinrib1983,Vojta2006,PastorSatorras2000}; but the
absence of any trend across $L=2048$--$8192$, the agreement between
independent estimators, and the coincidence of the exponents for
$\sigma_h=10$ and $\sigma_h=1$ make the \TFFM\ a well-defined target for
a theoretical treatment.  The natural starting point is a spreading process
in which each site carries a quenched, locally directed bias whose spatial
correlations are irrelevant at long wavelengths (the exponents do not depend
on $\sigma_h$) but whose asymmetry is not (the exponents differ from
isotropic percolation).

Two remarks on methodology follow from this analysis.  First, in models of
this type the finite-size width of the transition can be far smaller than
the natural $\beta$ resolution of a parameter sweep; a coarse grid then
produces smoothing-dependent and erratic estimates of $\nu$ that look like
slow crossover but are not.  Second, the infinite-randomness scenario that
such erratic estimates might suggest~\cite{Vojta2005} is disfavored here by
the time-dependent survival probability, which follows a power of $t$ rather
than of $\ln t$ and shows no $L$ dependence of its running exponent.  A scan
at $\sigma_h=3$ on the coarse grid ($L\leq2048$, $N=1000$) gives $\delta$ in
the range $0.32$--$0.36$, consistent with the other roughnesses; a fine-grid
determination of $\nu$ at intermediate $\sigma_h$, and the dependence of the
crossover to the smooth-terrain limit on $\sigma_h$, are natural next steps.

The thresholds $\betac^\infty$, the exponents $\nu$ and $\delta$, the
critical survival probability, and the terrain and wind phenomenology
reported below constitute the primary contributions of this work.

\subsection{Wind coupling: physical interpretation}
\label{sec:disc_wind}

The wind results of Sec.~\ref{sec:results_wind} establish three physically
significant findings that complement the terrain results.

\textit{Wind dominates over terrain in setting the fire risk threshold.}
Terrain coupling shifts $\betac$ by at most $\sim\!11\%$ over
$|\gamma|\in[0,3]$ at $\sigma_h=10$, $p=0.8$.  Wind coupling, by contrast,
raises $\betac$ by a factor of $\sim\!2$--$4$ across the studied range:
from $0.40$ (terrain-only) to $0.81$ (axial saturation) or $1.56$ (diagonal,
$\dw=2$) at $p=0.8$, and from $0.68$ to $2.05$ (axial) or $1.83$
(diagonal) at $p=1.0$.
In conditions representative of observed strong wind events, the suppression
effort required to contain a fire may be several times larger than the
terrain-only estimate---a quantitative prediction with direct implications for
resource allocation in fire management.

\textit{Clipping saturation.}
For axial wind at $p=0.8$, $\betac(\dw)$ saturates once downwind bonds reach
unit probability---a regime of wind-inefficiency absent from terrain-only models.
Beyond $\dw\approx 1$, additional wind intensity does not increase fire risk as
measured by $\betac$; in the limit $\dw\to\infty$ the model approaches directed
bond percolation~\cite{Hinrichsen2000}.

\textit{Drift as a sensitive onset indicator.}
The wind-direction drift $\avg{d_w}$ detects the onset of wind-driven behavior
more sensitively than $\betac$ or $\avg{\eta}$: a step from $\dw=0$ to
$\dw=0.25$ produces $\avg{d_w}\approx 310$--$355$ lattice units while
$\betac$ shifts by only $\approx15\%$.  Drift is measurable from satellite burn-scar data~\cite{Loboda2007}
and could serve as an observational proxy for wind coupling strength.

\textit{Terrain-wind competition and the wind paradox.}
A qualitatively distinct regime, with no counterpart in the \WFFM, emerges at
high terrain coupling ($\gamma\gtrsim 3$) on rough terrain: increasing east wind
\emph{decreases} the total burned fraction, inverting the conventional expectation
that stronger wind always enlarges fire.  At large $\gamma$, correlated terrain
gradients already bias fire along coherent pathways in all radial directions;
superimposing an east-wind bias funnels the propagating front into a narrow
downwind corridor that exits the eastern boundary before the terrain-driven radial
expansion can fill the domain, yielding a smaller but more elongated scar
at the moment the front reaches the boundary (Fig.~\ref{fig:app_paradox}).
In a model without spatially coherent bond pathways, such as the \WFFM, we
expect no such inversion: without terrain structure to compete against, wind
uniformly extends the fire.  The wind paradox thus provides a
model-discriminating observable between landscapes where terrain channeling and
wind cooperate, and landscapes where they compete; it may be identifiable in
satellite burn-scar records for events combining strong topographic relief with
sustained directional winds, and is absent from operational spread models that
treat terrain and wind as purely additive accelerators~\cite{Finney1998,Cruz2017}.

The burned fraction in Figs.~\ref{fig:app_morph} and~\ref{fig:app_paradox}
is evaluated when the front first reaches the boundary, so one might ask
whether the inversion is merely an artifact of the finite absorbing boundary.
A simple area argument indicates that it is not, and that the effect should
in fact become \emph{more pronounced} at larger $L$: terrain-driven radial
spread burns an $\mathcal{O}(1)$ fraction of the domain at boundary
crossing, since the scar area $\sim\pi(L/2)^2$ is a fixed fraction of $L^2$,
whereas wind-channeled spread burns a corridor of terrain-set width
$w\ll L$ before exiting the downwind boundary, contributing a burned
fraction $\sim w/L$ that decreases with $L$.  The physical content of the
effect is therefore geometric: at fixed elapsed time, or at fixed distance
traveled by the head of the fire, strong wind on strongly channeled terrain
produces a narrower scar.  A quantitative test at fixed burning time on
larger lattices is a natural extension.

Together, the risk-threshold amplification, the shape saturation, the drift onset,
and the terrain-wind paradox provide complementary diagnostics spanning the full
parameter space.  $\betac$ is most sensitive to strong wind; $\avg{\eta}$ saturates
early and signals the clipping regime; $\avg{d_w}$ detects weak-coupling onset;
and the decrease of burned fraction with wind signals the terrain-dominated regime.

\subsection{Implications for wildfire science}
\label{sec:disc_wildfire}

Our results carry several implications for the statistical description of real
wildfires.  The decrease of $\betac$ with $|\gamma|$ and with decreasing
$\sigma_h$ means that, in a model where the only terrain effect is slope
asymmetry of the spread probability, topography is a net obstacle to
large-scale percolation: fire climbs readily but stalls at ridgelines and
must find a way around descending flanks.  This is the lattice analog of the
well-documented role of ridgelines as natural firebreaks and of the
upslope--downslope asymmetry of spread rates~\cite{Sharples2009,WindSlope2026}.
The flip side is that the same landscapes are the ones in which wind matters
most: as shown in Sec.~\ref{sec:disc_wind}, a directional wind supplies the coherent bias that
terrain alone cannot, and the two together can either cooperate or compete.
The terrain correlation length $\sigma_h$ and slope coupling $\gamma$ are
directly extractable from digital elevation models and fuel maps, making
$\betac(\gamma,\sigma_h)$ a landscape-level metric of how much suppression
effort separates a fire from percolation.  We stress that real fires also
respond to terrain through fuel and moisture gradients and through
fire--atmosphere coupling~\cite{Sharples2009}, effects not included here
that can make slopes net accelerants.

Wind amplifies the fire risk threshold by a factor of $\sim\!2$--$4$ and
produces large downwind displacement of fire scars, consistent with the dominant
role of wind in observed catastrophic fire events~\cite{Cruz2017,Sharples2016,WindSlope2026}.
The rapid saturation of shape anisotropy ($\avg{\eta}\approx 0.85$ at
$\dw\gtrsim 0.75$) and the sharp onset of downwind drift suggest that both
observables are useful diagnostics in real-time fire monitoring.  The elongated,
downwind-displaced burn scars produced by the \TFFM\ with wind are directly
comparable to the geometry documented in empirical fire-spread studies in
plantation forests~\cite{DirectionalFire2025,ScientificReports2025} and in
the catastrophic Chilean fire seasons where strong Puelche winds repeatedly
drove fire across valley terrain into plantation monocultures~\cite{Bowman2019Chile,ChileDataset2025}.

We also note that the core physical picture is supported by observational
evidence: real fire scars exhibit fractal dimensions consistent with critical
percolation~\cite{Caldarelli2001}, anisotropic spread under terrain and wind
has been related to directed-percolation universality~\cite{Porterie2008},
and the February~2024 Valpara\'iso fire in Chile---in which fire channeled
through valley corridors under strong winds and grew several-fold within
hours~\cite{NASAValpo2024}---provides a qualitative realization of the
strongly channeled, wind-driven regime in which our model predicts elongated
scars and large downwind drift.  Whether the competitive (burned-area
reducing) regime identified here can be recognized in burn-scar records is
a testable question that operational models treating terrain and wind as
purely additive accelerators~\cite{Finney1998,Cruz2017} would not raise.

The \TFFM\ with wind coupling remains intentionally minimal: it does not include
moisture dynamics, species-level fuel variability, or fire--atmosphere feedback.
These omissions limit direct quantitative comparison to specific fire events.
Nonetheless, the model's ability to reproduce qualitatively realistic fire
shapes, a monotone phase boundary from a small number of parameters, and the
sharp contrast between terrain and wind contributions suggest that this
framework captures essential features of the active-to-inactive fire
transition in heterogeneous, wind-exposed landscapes.

\section{Conclusions}
\label{sec:conclusions}

We have introduced the \TFFM\ and characterized its phase structure, critical
behavior, and response to wind.  Our main findings are as follows.

\textit{Phase boundary and terrain symmetry.}
$\betac$ decreases monotonically with $|\gamma|$, falling by $\sim\!11\%$ over
$|\gamma|\in[0,3]$ at $p=0.8$, and satisfies $\betac(\gamma)\approx\betac(-\gamma)$
to numerical precision.

\textit{Monotone $\betac(\sigma_h)$ and the always-inactive regime.}
$\betac(\sigma_h)$ decreases monotonically as $\sigma_h$ decreases: rough
terrain shrinks the active phase because slope asymmetry penalizes downhill
bonds and traps fire at local elevation maxima, whose density grows as
$\sigma_h^{-2}$.  For sufficiently small $\sigma_h$ and low $p$, $\betac$
falls below zero and no fire percolates even at zero suppression
($\betac<0$ at $p\leq 0.8$ for $\sigma_h\leq 0.5$, and at $p=0.7$ for
$\sigma_h\leq 1$).

\textit{Shape anisotropy and fire-scar geometry.}
$\avg{\eta}$ rises steeply at $\betac$ and remains large in the inactive
phase; its near-critical rise is steeper and higher on smooth terrain
($\sigma_h\gtrsim 10$), which sustains coherent directional channels that
produce elongated burn scars.

\textit{Finite-size scaling.}
$\betac(L)$ shifts monotonically with $L$ for both $\sigma_h$ values, with a
total shift $\Delta\betac\approx0.012$ from $L=256$ to $L=8192$.  The
thermodynamic critical thresholds are $\betac^\infty=0.398\pm0.001$
($\sigma_h=10$) and $0.188\pm0.001$ ($\sigma_h=1$).  On a fine $\beta$
grid at $L=2048$--$8192$ the susceptibility peak and the width of the sharp
component of $\psurviv$ give a correlation-length exponent $\nu=1.8\pm0.2$
for both roughnesses, and the velocity exponent is $\delta=0.34\pm0.03$,
stable for $L\geq1024$.  Neither exponent matches directed percolation
($\nu_\perp=0.733$; front-velocity exponent $\nu_\parallel-\nu_\perp\approx0.56$),
isotropic percolation ($\nu=4/3$; $\nu(d_{\min}-1)\approx0.18$), or compact
directed percolation ($\nu_\perp=1$).  Coarse-grid estimates of $\nu$ are
shown to be resolution artifacts, a caution relevant to other
quenched-disorder spreading models.

\textit{Survival probability at criticality.}
The single-seed survival probability $P(t)$ at $\betac^\infty$ is independent
of $L$ up to $8192$ and decays extremely slowly, from $0.85$ at $t=10$ to
$0.52$ at $t\approx8000$, with a running exponent that falls from
$\approx0.09$ to $0.03$--$0.05$; directed percolation ($0.451$) is excluded
directly, and the data are best described by an approach to a finite
$P_\infty\approx0.3$--$0.5$.  Together with the $L$-independent jump of
$\psurviv$ from $P^*\approx0.5$ to zero and the two-component transition
profile resolved on a fine $\beta$ grid, this points to a transition at
which the survival probability is discontinuous while the front velocity
vanishes continuously, $\avg{v}\sim(\betac-\beta)^{0.34}$.

\textit{Wind coupling.}
Wind raises $\betac$ by a factor of $\sim\!2$--$4$ across the studied
range, far exceeding the terrain contribution alone.  For axial wind at
$p=0.8$, $\betac(\dw)$ saturates once downwind bonds are clipped to unit
probability---a new regime absent in the terrain-only model.  Shape anisotropy
saturates rapidly at $\avg{\eta}\approx 0.85$ for $\dw\gtrsim 0.75$,
independently of wind direction.  The wind-direction drift $\avg{d_w}$
exhibits a sharp onset at $\dw=0.25$ ($\avg{d_w}\approx 310$--$400$ lattice
units) followed by a plateau, making it the most sensitive observable for
detecting the onset of wind-driven fire behavior.

\textit{Terrain-wind competition.}
At high terrain coupling ($\gamma\gtrsim 3$) on rough terrain, east wind
\emph{decreases} the burned fraction at boundary crossing by funneling fire
into a narrow downwind corridor that exits the domain before the
terrain-driven radial expansion fills it; this inversion of the usual
wind enhancement has no analog in isotropic bond-disorder models and emerges
directly from the competition between coherent terrain pathways and wind
directionality.  A simple area argument indicates that the inversion
should become more pronounced with increasing $L$, since the terrain-driven
burned fraction is $\mathcal{O}(1)$ while the wind-channeled fraction scales
as $w/L$ with a terrain-set corridor width $w$.  The effect provides a
model-discriminating observable between cooperative and competitive
terrain-wind regimes (Figs.~\ref{fig:app_morph} and~\ref{fig:app_paradox}).

The observables introduced here---shape anisotropy, fire-front roughness,
wind-direction drift, and the terrain-wind inversion---are directly accessible
from satellite-derived burn-scar databases and provide quantitative targets
for model validation.  Natural extensions include coupling tree density and fuel moisture to
terrain elevation, so that slopes can act as net accelerants as well as
barriers, studying the interaction of terrain and wind
anisotropies~\cite{Sharples2016}, examining non-Gaussian terrain models
(e.g., fractal landscapes), and resolving the asymptotic finite-size scaling
of the transition with derivative-free estimators of $\nu$.

\begin{acknowledgments}
J.M.F. and C.M. thank Universidad Central de Chile.
E.S.M. acknowledges financial support from Proyecto Interno USM 2026 PI\_LIR\_26\_12.
J.R. thanks “Estrategia de Sostenibilidad del Grupo de Magnetismo y Simulación de la Universidad de Antioquia a través del proyecto con código ES84230022”.
\end{acknowledgments}

\section*{AI-use disclosure}
AI assistants (Claude models, Anthropic) were used to assist with manuscript
editing, the development of parallelized simulation code, figure generation,
and bibliographic checking.  All simulations were performed on the
computational resources of the Grupo de Simulaciones, Departamento de
F\'isica, UTFSM.  All AI-assisted output was reviewed by the authors, who
designed the model and the numerical experiments, verified all physical
content and numerical results, and take full responsibility for the final
text.

\appendix

\section{Tabulated Critical Thresholds}
\label{app:tables}

Tables~\ref{tab:betac_gamma} and \ref{tab:betac_sigmah} report the numerically
measured critical thresholds $\betac$ for all simulated parameter combinations.
All values were extracted from the $\psurviv(\beta)=0.5$ crossing at
$L=2048$, $N=2000$.  The dagger ($\dagger$) at $(p=1.0,\sigma_h=100)$
indicates that the crossing was not bracketed within the scanned range;
entries showing $\ldots$ indicate parameter combinations not simulated
(Table~\ref{tab:betac_gamma}) or those for which $\betac<0$, i.e., the
always-inactive regime with $\psurviv\equiv0$ (Table~\ref{tab:betac_sigmah}).

\begin{table}[htb]
\caption{\label{tab:betac_gamma}%
  $\betac$ ($\psurviv=0.5$ crossing) vs $\gamma$ at $\sigma_h=10$,
  $L=2048$, $N=2000$. Ellipses: $\gamma$ not simulated.
  Even symmetry $\betac(\gamma)\approx\betac(-\gamma)$ holds to within
  the stated precision.}
\begin{ruledtabular}
\begin{tabular}{rcccc}
$\gamma$ & $p=0.7$ & $p=0.8$ & $p=0.9$ & $p=1.0$ \\
$-2.0$            & 0.2027 & 0.3830 & 0.5355 & \ldots \\
$-1.0$            & 0.2211 & 0.3987 & 0.5516 & 0.6834 \\
$-0.5$            & 0.2263 & 0.4029 & 0.5529 & \ldots \\
$\phantom{-}0.0$  & 0.2274 & 0.4047 & 0.5573 & 0.6918 \\
$\phantom{-}0.5$  & 0.2259 & 0.4033 & 0.5562 & 0.6901 \\
$\phantom{-}1.0$  & 0.2213 & 0.3984 & 0.5516 & 0.6836 \\
$\phantom{-}1.5$  & 0.2140 & 0.3919 & 0.5448 & \ldots \\
$\phantom{-}2.0$  & 0.2037 & 0.3837 & 0.5362 & 0.6683 \\
$\phantom{-}3.0$  & 0.1737 & 0.3609 & 0.5151 & \ldots \\
\end{tabular}
\end{ruledtabular}
\end{table}

\begin{table}[htb]
\caption{\label{tab:betac_sigmah}%
  $\betac$ vs $\sigma_h$ at $\gamma=1.0$, $L=2048$, $N=2000$.
  Ellipses: $\psurviv=0.5$ crossing not found in $\beta\in[0,1.1]$,
  i.e., $\betac<0$ (always-inactive regime, $\psurviv\equiv0$).
  $\betac$ decreases monotonically as $\sigma_h\to 0$.}
\begin{ruledtabular}
\begin{tabular}{rcccc}
$\sigma_h$ & $p=0.7$ & $p=0.8$ & $p=0.9$ & $p=1.0$ \\
$0$    & \ldots & \ldots & \ldots & 0.2867 \\
$0.5$  & \ldots & \ldots & 0.2011 & 0.3935 \\
$1$    & \ldots & 0.1886 & 0.3891 & 0.5468 \\
$1.5$  & 0.0498 & 0.2907 & 0.4651 & 0.6086 \\
$2$    & 0.1185 & 0.3339 & 0.4981 & 0.6345 \\
$3$    & 0.1779 & 0.3687 & 0.5239 & 0.6586 \\
$5$    & 0.2090 & 0.3889 & 0.5406 & 0.6754 \\
$10$   & 0.2213 & 0.3984 & 0.5516 & 0.6836 \\
$20$   & 0.2247 & 0.4020 & 0.5559 & 0.6901 \\
$50$   & 0.2265 & 0.4044 & 0.5579 & 0.6912 \\
$100$  & 0.2257 & 0.4053 & 0.5550 & $\dagger$ \\
\end{tabular}
\end{ruledtabular}
\end{table}

\section{Terrain geometry and spreading-probability visualizations}
\label{app:figs}

\begin{figure*}[t]
\includegraphics[width=\textwidth]{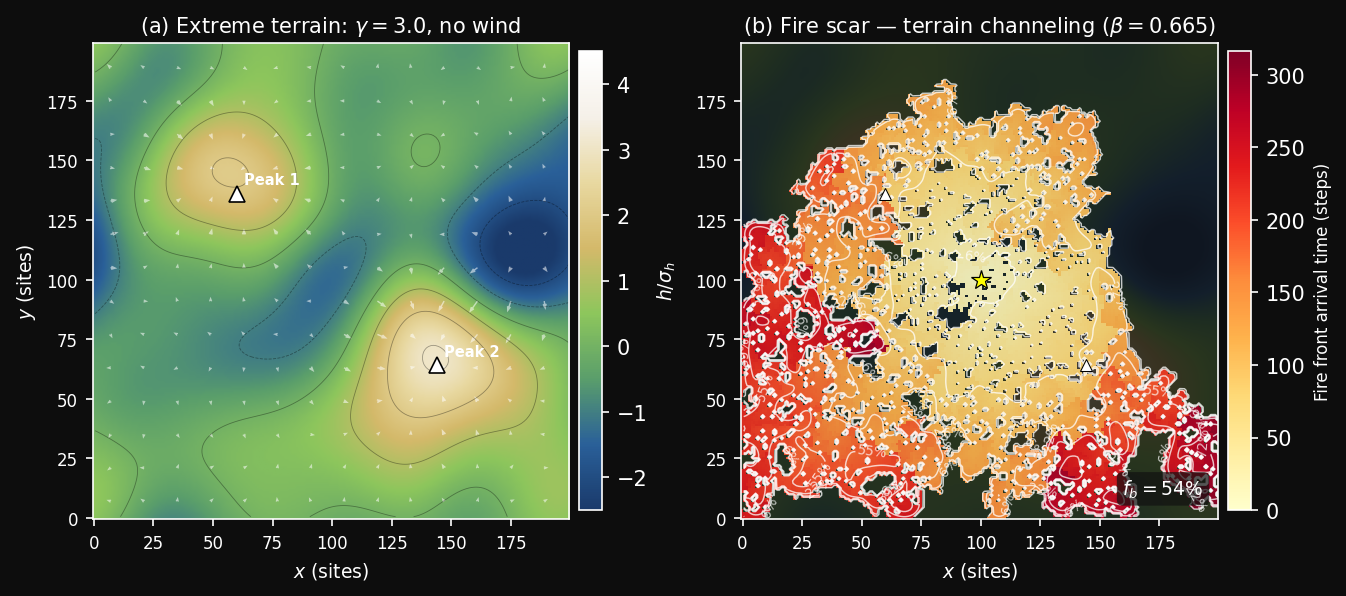}
\caption{%
  \textbf{Extreme terrain: two dominant peaks channel fire propagation.}
  \textbf{(a)} Terrain ($L=200$): Gaussian background ($\sigma_h=18$) with
  two superimposed peaks (amplitudes $5.5\,\sigma$, $5.0\,\sigma$;
  half-width $\approx28$ sites; white triangles). Arrows: local uphill gradient.
  \textbf{(b)} Fire scar ($\gamma=3.0$, $\beta=0.665$, $p=1.0$, no wind);
  color = front arrival time (light = early, dark red = late); white contours
  = isochrones; yellow star = ignition. The scar ($f_b\approx54\%$) engulfs
  both peaks and the saddle between them; the descending flanks into the
  surrounding basins (blue in panel a) act as barriers because downhill bonds
  are suppressed for $\gamma>0$.}
\label{fig:app_ext}
\end{figure*}

\begin{figure*}[t]
\includegraphics[width=\textwidth]{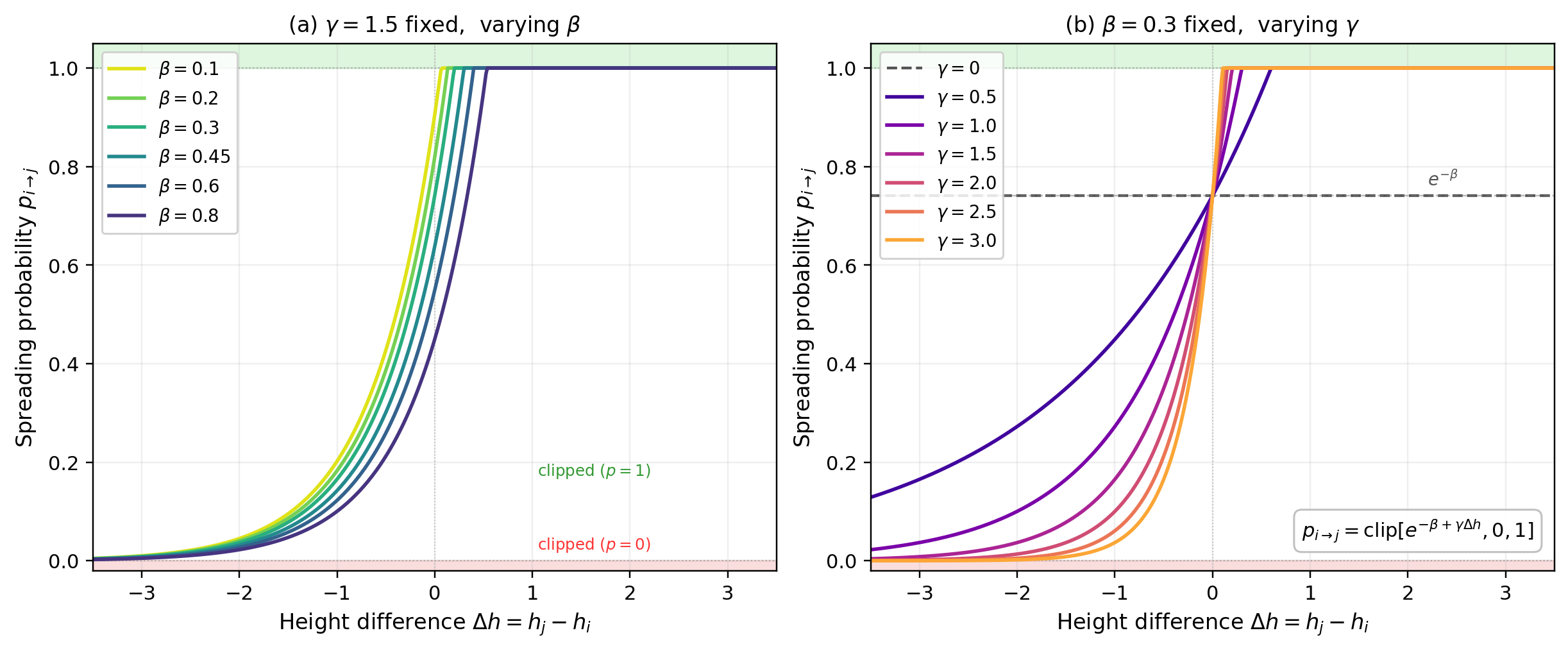}
\caption{%
  \textbf{Spreading probability $p_{i\to j}$ vs height difference.}
  Left: $p_{i\to j}(\Delta h)$ for $\gamma=1.5$ and
  $\beta\in\{0.1,0.2,0.3,0.45,0.6,0.8\}$.
  Right: $p_{i\to j}(\Delta h)$ for $\beta=0.3$ and
  $\gamma\in\{0,0.5,1.0,1.5,2.0,2.5,3.0\}$; the dashed line is the flat-terrain
  value $e^{-\beta}$.
  Clipping to $[0,1]$ creates a plateau at unity for uphill bonds
  ($\Delta h=h_j-h_i>0$, $\gamma>0$) and exponentially suppresses downhill
  spreading. $\beta$ and $\gamma$ together control the width and asymmetry
  of the active-spreading window.}
\label{fig:app_pij}
\end{figure*}

Figures~\ref{fig:app_ext} and~\ref{fig:app_pij} provide additional visualizations
of the \TFFM\ terrain geometry and spreading-probability function that supplement
the main schematic (Fig.~\ref{fig:schematic}).  Figure~\ref{fig:app_ext} shows
an extreme-terrain scenario dominated by two large-amplitude peaks: with
$\gamma>0$ the fire ignited between them climbs both peaks and the saddle that
connects them, while the descending flanks that lead into the surrounding
basins, where every outgoing bond is penalized, act as barriers and remain
unburned.  The model thus captures topographic channeling even when the
landscape is dominated by a small number of large features.

\section{Fire-scar parameter sensitivity}
\label{app:application}

\begin{figure*}[t]
\includegraphics[width=\textwidth]{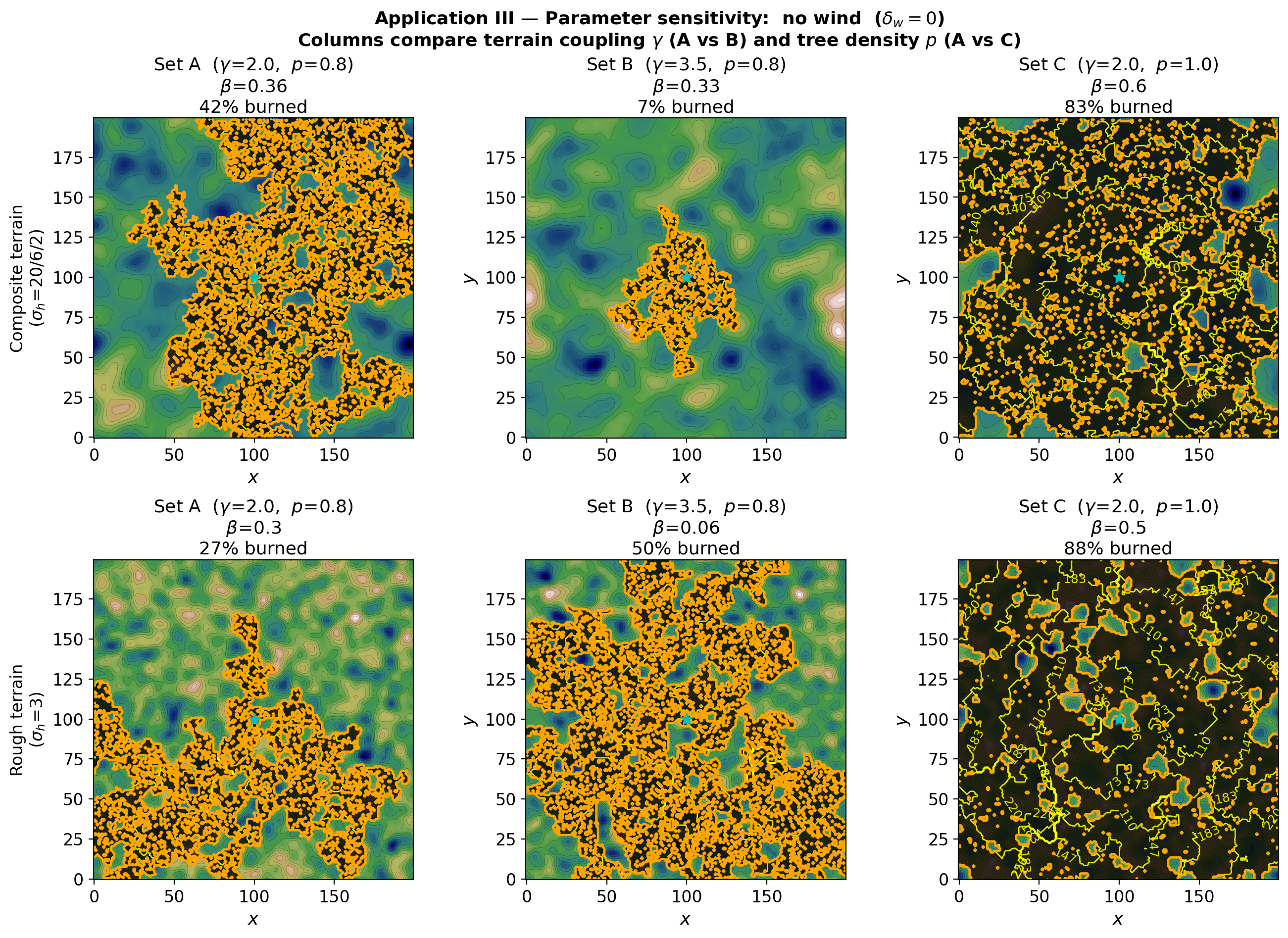}
\caption{%
  \textbf{Parameter sensitivity: $\gamma$ and $p$ at $\dw=0$.}
  Top: composite terrain; bottom: rough terrain.
  Columns: Set~A ($\gamma=2.0$, $p=0.8$), Set~B ($\gamma=3.5$, $p=0.8$),
  Set~C ($\gamma=2.0$, $p=1.0$), each near its own $\betac$.
  Set~B scars follow elevation contours more tightly (stronger $e^{\gamma\Delta h}$);
  Set~C ($p=1.0$) shows fewer interior micro-patches, isolating the
  site-dilution contribution from bond-disorder and terrain coupling.
}
\label{fig:app_params}
\end{figure*}

Figure~\ref{fig:app_params} compares Sets~A, B, and C at $\dw=0$ on both
terrain types, isolating the roles of $\gamma$ and $p$ independently.
Each panel shows a single near-critical realization at $L=200$; yellow isochrones
mark equal-time fronts, the orange curve delimits the scar, and the cyan star
marks the ignition site.  Set~B ($\gamma=3.5$) produces more irregular,
terrain-following scars than Set~A ($\gamma=2.0$) on the same substrate, as the
stronger terrain factor $e^{\gamma\Delta h}$ amplifies elevation differences and
channels fire more sharply along gradient lines.  Set~C ($p=1.0$) eliminates the
site-percolation component, yielding scars with noticeably fewer interior
micro-patches while boundary morphology remains governed by bond disorder and
terrain coupling, directly isolating how $p$ modulates fire-scar texture
independently of the spreading rule.  The terrain-type and wind-coupling effects
are shown in full in Figs.~\ref{fig:app_morph} and~\ref{fig:app_paradox} of the
main text.

\bibliography{tffm_references}

\end{document}